\documentclass[journal]{IEEEtran}
\usepackage{amssymb,amsmath}
\usepackage{booktabs}
\usepackage{siunitx}

\ifCLASSINFOpdf
\else
   \usepackage[dvips]{graphicx}
\fi
\usepackage{url}

\usepackage{hyperref}
\usepackage{csquotes}
\usepackage{graphicx}
\usepackage{comment}
\usepackage[margin=0.7in]{geometry}
\usepackage{pgfplots}
\pgfplotsset{compat=1.17}
\usepackage{pgfplotstable}
\usepackage{pgfplotstable}\usepackage{multirow}
\usepackage{pgfplots}
\pgfplotsset{compat=newest}
\usetikzlibrary{patterns, shapes, arrows}
\usepackage[linesnumbered,ruled,vlined]{algorithm2e}
\usepackage{bm}
\usepackage{textcomp}
\usepackage{stfloats}
\usepackage{url}
\usepackage{verbatim}
\usepackage{graphicx}
\usepackage{cite}
\usepackage{csquotes}
\usepackage{enumitem}

\usepackage{hyperref}
\usepackage{pifont}
\usepackage{graphicx}
\usepackage{subcaption}
\usepackage{mathtools}
\usepackage[table]{xcolor}
\usepackage{booktabs}
\usepackage{arydshln}
\usepackage{float}
\begin{document}

\title{Compressing Quaternion Convolutional Neural Networks for Audio Classification}
\author{Arshdeep Singh, Vinayak Abrol, \IEEEmembership{Senior Member, IEEE} and Mark D. Plumbley \IEEEmembership{Fellow, IEEE}
\thanks{This work was done while Arshdeep Singh and Mark D. Plumbley were at Centre for Vision, Speech and Signal Processing, University of Surrey, U.K. Arshdeep Singh and Mark D. Plumbley are now with King's College London, U.K. (e-mail: arshdeep.singh@kcl.ac.uk; mark.plumbley@kcl.ac.uk). Vinayak Abrol is at  Infosys Centre for AI and CSE Department, Indraprastha Institute of Information Technology (IIIT) Delhi, India (e-mail: abrol@iiitd.ac.in)}
\thanks{}}

\markboth{Journal of \LaTeX\ Class Files, Vol. 14, No. 8, August 2015}
{Shell \MakeLowercase{\textit{et al.}}: Bare Demo of IEEEtran.cls for IEEE Journals}
\maketitle

\begin{abstract}

Conventional Convolutional Neural Networks (CNNs) in the real domain have been widely used for audio classification. However, their convolution operations process multi-channel inputs independently, limiting the ability to capture correlations among channels. This can lead to suboptimal feature learning, particularly for complex audio patterns such as multi-channel spectrogram representations. Quaternion Convolutional Neural Networks (QCNNs) address this limitation by employing quaternion algebra to jointly capture inter-channel dependencies, enabling more compact models with fewer learnable parameters while better exploiting the multi-dimensional nature of audio signals. However, QCNNs exhibit higher computational complexity due to the overhead of quaternion operations, resulting in increased inference latency and reduced efficiency compared to conventional  CNNs, posing challenges for deployment on resource-constrained platforms. To address this challenge, this study explores knowledge distillation (KD) and pruning, to reduce the computational complexity of QCNNs while maintaining performance. Our experiments on audio classification reveal that pruning QCNNs achieves similar or superior performance compared to KD while requiring less computational effort. Compared to conventional CNNs and Transformer-based architectures, pruned QCNNs achieve competitive performance with a reduced learnable parameter count and computational complexity. On the AudioSet dataset, pruned QCNNs reduce computational cost by 50\% and parameter count by 80\%, while maintaining performance comparable to the conventional CNNs. Furthermore, pruned QCNNs generalize well across multiple audio classification benchmarks, including GTZAN for music genre recognition,  ESC-50 for environmental sound classification and RAVDESS for speech emotion recognition. 


\end{abstract}

\begin{IEEEkeywords}
Computational complexity, CNNs, Quaternions, Sustainable AI, AudioSet, Music Genre classification, RAVDESS, ESC-50.
\end{IEEEkeywords}

\IEEEpeerreviewmaketitle

\section{Introduction}

\IEEEPARstart{C}onvolutional neural networks (CNNs) \cite{mesaros2019sound, kong2020panns} are widely used in audio classification tasks due to their architectural properties such as locality and weight sharing. These architectural inductive biases facilitate learning from limited data and contribute to computational efficiency \cite{wang2023theoretical}, which is beneficial for real-time and embedded applications.  More recently, Transformer-based architectures \cite{gong2021ast, chong2023masked} have gained attention for their ability to capture long-range dependencies through self-attention mechanisms \cite{vaswani2017attention, d2021convit}. However, the lack of built-in inductive priors in Transformers typically requires larger datasets and greater computational resources during training \cite{goyal2022inductive}. While Transformer models are increasingly adopted in the field, CNNs remain relevant and continue to be used widely in audio classification, particularly in data- and resource-constrained settings \cite{mesaros2025decade}. Therefore, this paper focuses on CNNs for audio classification.

In conventional real-valued  CNNs, sharing of learnable parameters (weights or convolution filters) is primarily applied across spatial dimensions of multiple channels including input channels and output channels (produced after convolution operation), enabling  CNNs to capture translation-invariant features within each channel. However, the convolution operation processes the channels independently using separate sets of parameters  for each channel, and subsequently aggregates them through summation (see Equation 1 in the \textcolor{black}{Supplemental} material) \cite{zhu2018quaternion}. This design results in limited or indirect inter-channel parameter sharing, which may limit the CNN’s capacity to explicitly capture internal dependencies between channels \cite{zhu2018quaternion, parcollet2018quaternion}.

Complex-valued CNNs provide an alternative to conventional CNNs, offering an efficient way to process multi-channel data \cite{parcollet2018quaternion}. In conventional CNNs, complex-valued inputs are typically separated into real and imaginary parts and processed as independent real-valued multiple channels \cite{trabelsi2018deep}.
On the other hand, complex-valued CNNs are able to capture both magnitude and phase relationships in a unified manner by preserving the intrinsic structure of complex data using convolution operation that uses complex multiplication \cite{yadav2023fccns}. Prior studies \cite{amin2011complex, nitta2004orthogonality} have demonstrated that complex-valued neural networks require fewer learnable parameters for function approximation and exhibit improved generalization capabilities over conventional real-valued networks.

While complex-valued CNNs capture interactions between two channels,  quaternion-valued CNNs (QCNNs) naturally extend this capability to four channels, incorporating one real and three imaginary components, with convolution operations involving the Hamilton  product \cite{parcollet2020survey}.  Eleonora et al. \cite{quaternion_complex} and Gaudet et al. \cite{gaudet2018deep} found that QCNNs outperform  complex-valued networks while using fewer learnable parameters, without significantly increasing the computational cost.
However, the Hamilton product based convolution operations on quaternion-valued variables have significantly higher computational complexity compared to that of scalar multiplication based convolutions on real-valued inputs in conventional CNNs \cite{parcollet2018quaternion,gaudet2018deep,zhang2019quaternion}.  For instance, each quaternion Hamilton product  consists of 16 scalar multiplications and 12 scalar additions.
Consequently, the lower parameter count of quaternion CNNs comes with the trade-off of increased computational complexity.

In this paper, we address the computational complexity associated with QCNNs and further improve their efficiency by reducing both  the number of learnable parameters and the computation complexity for audio classification.  We explore the use of QCNNs for  audio classification tasks, such as audio tagging, music genre classification, environmental sound classification and  speech emotion recognition. QCNNs are relatively less explored for audio compared to the advancements seen in the image processing domain \cite{parcollet2020survey,miron2023quaternions}.

This work extends our previous work \cite{chaudhary2024efficient} on reducing the size of memory storage and the  computation complexity of convolutional neural networks. In that work, we improved the efficiency of a  conventional real-valued pre-trained CNN14 \cite{kong2020panns}, convolutional neural network designed for an audio tagging task, by transforming it into quaternion network and then applying a $l_1$-norm-based filter pruning method \cite{li2017pruning} to reduce the number of computations and the memory required.


In this paper, we reduce the number of computations of  residual quaternion convolutional networks such as a quaternion ResNet38 and feedforward quaternion convolutional networks without any  residual connections, such as a quaternion CNN14, using different model compression methods including filter pruning and knowledge distillation.
We also analyse the generalization performance of the compressed QCNNs across different classification tasks, including music genre classification, environmental sound classification and  speech emotion recognition. We also compare the environmental impact of compressed and original  QCNNs. 



The key contributions of the paper are summarized as follows:
\begin{itemize}
    \item We propose a model compression pipeline to enhance the efficiency of Quaternion Convolutional Neural Networks (QCNNs) for audio classification.

    \item We explore model compression techniques, including  knowledge distillation (KD) and filter pruning, to reduce computational costs and memory requirements for both feedforward and residual QCNNs.

    \item We find that pruning methods generally outperform KD, and that pruning QCNNs offers a better trade-off between accuracy and resource efficiency compared to pruning conventional CNNs.
    
    \item We find that compared to existing CNNs and Transformer  models  pruned QCNNs show similar or better performance while using significantly fewer parameters across multiple audio classification benchmarks. These include GTZAN \cite{tzanetakis2002musical} for music genre recognition, ESC-50 \cite{piczak2015esc} for environmental sound classification and RAVDESS \cite{livingstone2012ravdess} for speech emotion recognition. \item Pruning QCNNs  reduces their environmental impact compared to  that of original QCNNs. 

\end{itemize}
    Our codes are available at Github\footnote{\href{https://github.com/Arshdeep-Singh-Boparai/audio-pruning.git}{https://github.com/Arshdeep-Singh-Boparai/audio-pruning.git}}.

\section{A background on Quaternion and conventional to quaternion CNN conversion}


A quaternion is typically defined as \cite{zhu2018quaternion}

\[
\mathbf{q} = q_R + q_I\boldsymbol{i} + q_J\boldsymbol{j} + q_K \boldsymbol{k},
\]

\noindent where  \( q_R, q_I, q_J, q_K \in \mathbb{R} \) are four scalar parameters, $q_R$ is the real component, $(\boldsymbol{i}, \boldsymbol{j}, \boldsymbol{k})$ are the orthonormal basis elements and denote the three imaginary axis, and ($q_I$, $q_J$, $q_K$) denote the three imaginary components.

The Hamilton product ($\otimes$) between two quaternions 
$\mathbf{m} = m_R + m_I\boldsymbol{i} + m_J\boldsymbol{j} + m_K\boldsymbol{k}$, and $\mathbf{q}$  is expressed as \cite{zhu2018quaternion}


\begin{equation}
\mathbf{m} \otimes \mathbf{q} =
\begin{bmatrix}
m_R & -m_I & -m_J & -m_K \\
m_I &  m_R & -m_K &  m_J \\
m_J &  m_K &  m_R & -m_I \\
m_K & -m_J &  m_I &  m_R \\
\end{bmatrix}
\begin{bmatrix}
q_R \\
q_I \\
q_J \\
q_K \\
\end{bmatrix}.
\end{equation}


In quaternion CNNs, the input is typically a multi-channel signal represented in quaternion form. The input and output channels in convolutional layers are also quaternion-valued, where each channel corresponds to a quaternion feature map produced by convolution operation between quaternion-valued input and quaternion-valued convolutional filter using the Hamilton product \cite{parcollet2018quaternion}.
A single quaternion-valued filter is shared across all input channels, and the use of the Hamilton product allows the filter to capture input inter-channel dependencies directly, without the need for additional parameters unlike real-valued convolution in conventional CNNs \cite{gaudet2018deep, zhang2019quaternion}. Further details on the Hamilton product and convolution in QCNNs  can be found in Section I of the \textcolor{black}{supplemental} material.





\subsection{Conventional to quaternion CNN architecture conversion}



\label{sec: quaternion model conversion}

Designing a quaternion variant of a conventional CNN requires 
(1) replacing the standard matrix-vector product with the Hamilton product to preserve quaternion interactions, and (2) adopting a split activation function, where the activation function is applied independently to each component of the quaternion \cite{parcollet2020survey}. Other components required to convert a conventional CNN to quaternion CNN include using quaternion batch normalization, quaternion fully-connected layer and quaternion pooling layer. A detailed overview about these components can be found in \cite{gaudet2018deep, altamirano2024quaternion}.



Furthermore, the input and output channels in convolutional layers of conventional CNNs are reorganized into groups of four channels to represent each group as a quaternion channel.

\begin{figure*}[t]
    \centering
    \includegraphics[width = \textwidth]{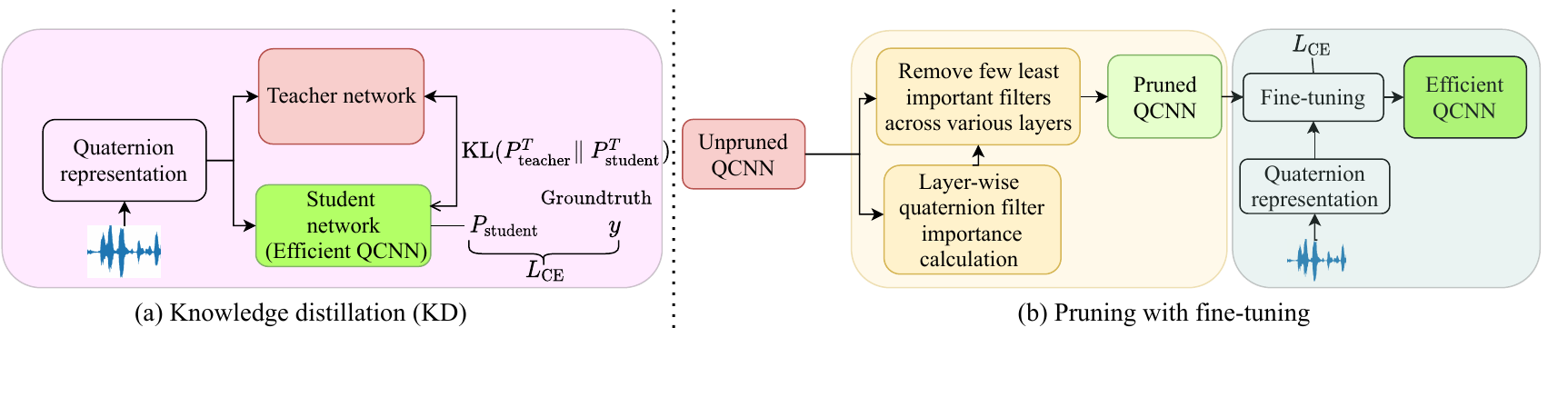}
    \vspace{-1.25cm}
    \caption{Quaternion CNN compression pipeline:(a) Knowledge distillation and (b) Pruning with fine-tuning.}
    \label{fig: overall framework}
\end{figure*}

\section{Model compression}
Model compression \cite{dantas2024comprehensive} aims to reduce the number of learnable parameters and the computational complexity of large-scale models without significantly compromising their performance. This is useful for deploying models on resource-constrained devices such as mobile phones, embedded systems, and IoT devices. Commonly used model compression methods are knowledge distillation and pruning.

\textbf{Knowledge distillation (KD)} is a widely used  model compression technique which leverages knowledge learned from 
a high-capacity teacher model to distil the knowledge in a smaller student model. KD allows the student to mimic the behaviour of the teacher, often achieving competitive accuracy with reduced computational and storage requirements \cite{gou2021knowledge}.

In KD, the student  and teacher models are aligned by minimizing  a loss such as the Kullback–Leibler (KL) divergence loss between  softened output probabilities generated by the teacher model ($P^T_{\text{teacher}}$) and the student model ($P^T_{\text{student}})$ ~\cite{hinton2015distilling}. The softened outputs are obtained by applying a softmax function with an elevated temperature 
\(T > 1\) to the unscaled output of model before applying an activation function, 
as given in Equation \ref{equ: softened output}.

\begin{equation}
\begin{aligned}
P^T_{\text{student}} &= \text{softmax}\!\left(\tfrac{z_{\text{student}}}{T}\right), \\
P^T_{\text{teacher}} &= \text{softmax}\!\left(\tfrac{z_{\text{teacher}}}{T}\right),
\end{aligned}
\label{equ: softened output}
\end{equation}

\noindent where $z_{\text{student}}$ and $z_{\text{teacher}}$ are the outputs generated by the student and the teacher model.





\textbf{Pruning} is a model compression method where learned parameters such as individual weights or entire convolutional filters  in the convolutional layers of convolutional neural networks \cite{he2023structured} are eliminated based on their \enquote{importance} \cite{he2023structured}, defined to mean how well the learned parameters contribute to the overall performance of the network.

Eliminating individual weights from convolutional layer result in an unstructured sparse pruned network. Practical acceleration of such unstructured sparse pruned networks is limited due to the random connections \cite{luo2018thinet}. Moreover, the unstructured sparse networks are not supported by off-the-shelf libraries and require specialised software or hardware for speed-up \cite{liang2021pruning}. 
On the other hand, filter pruning eliminates whole filter from the convolutional layer that result in a structured pruned network \cite{he2023structured}  and does not require additional resources for speed-up.


In filter pruning, the importance of convolutional filters is estimated using either data-dependent methods~\cite{li2024efficient, luo2020autopruner} or data-independent methods~\cite{li2017pruning, he2019filter}. Data-dependent methods rely on the training dataset to generate output from the filters, from which statistical measures, such as the average percentage of zeros or similarity between channels, are used to quantify filter importance. 
In contrast, data-independent methods, such as those based on the $l_1$-norm~\cite{li2017pruning} or the geometric-median~\cite{he2019filter} or the operator-norm \cite{singh_transaction} evaluate the filter importance using only the filter parameters. 
Once the importance of the filters is obtained,  a user-defined pruning ratio is used to eliminate the least important filters and the remaining filters from the original or unpruned model are retained.

In this paper, following our previous work \cite{chaudhary2024efficient}, we use  data-independent filter pruning methods to obtain a pruned model. To regain perform loss due to pruning, we then fine-tuned the pruned model on the original dataset until the validation accuracy stabilizes or improves, enabling the pruned model to recover relevant representations affected by pruning. 



\section{Model compression for Quaternion CNNs}
\label{sec: quaternion CNN compression}

In this section, we discuss the steps to compress QCNNs.


\subsection{QCNN model compression via knowledge distillation}

In this method, we build a small-size 
QCNN architecture and then use it as the student model,  trained using the KD method from the  pre-trained teacher model.  KD uses two loss functions: the cross-entropy loss ($\text{L}_{\text{CE}}$) and the distillation loss (KL divergence)  as shown in Figure \ref{fig: overall framework}(a). The total loss is the weighted sum of the two losses and is given below,

\begin{equation}
L_{\text{total}} = \alpha \cdot L_{\text{CE}}\big(y, P_{\text{student}}\big) 
+ (1 - \alpha) \cdot \text{KL}~\!\big(P_{\text{teacher}}^{T} \,\|\, P_{\text{student}}^{T}\big),
\label{equ: kd_total}
\end{equation}

\noindent where $L_{\text{CE}}$ is the cross-entropy loss between the one-hot groundtruth label $y$ and the student's predicted probability distribution  $P_{\text{student}}=\text{softmax}(z_{\text{student}})$ and the $\text{KL}$ divergence loss is between the softened teacher distribution $P_{\text{teacher}}^{T}$  and the softened student distribution $P_{\text{student}}^{T}$. The $\alpha \in [0,1]$ is a hyperparameter.



\subsection{QCNN model compression via  data-independent filter pruning}
\label{sec: filter pruning}

The filter pruning method consists of two steps. Firstly, we evaluate the importance of each quaternion filter in the unpruned QCNN layer-by-layer. 
The importance of quaternion filters is computed using one of the three methods: (a) the $l_1$-norm \cite{li2017pruning}, (b) the geometric-median \cite{he2019filter} and (c) the operator-norm \cite{singh_transaction}.


Secondly, we prune (remove) the $p$ least important quaternion filters from convolutional layers that contribute most significantly to the total parameter count of the network. The pruning ratio $p$ is a hyperparameter and is user-specified.


There will be some performance loss introduced by pruning. To reduce this,  
the remaining filters are then fine-tuned on the original dataset until the pruned model recovers the performance affected by pruning. 
A summary of quaternion filter pruning pipeline is given in Algorithm \ref{alg: filter pruning} and is shown in Figure \ref{fig: overall framework}(b).

Now, we describe various pruning methods used to compute importance of quaternion-valued filters in QCNNs.

Let us denote that the $m^{th}$ quaternion filter in the $l^{th}$ layer of QCNN as $\mathbf{F}^{l,m}$ =  $\mathbf{F}_R^{l,m}$ + $\mathbf{F}_I^{l,m}\boldsymbol{i}$ + $\mathbf{F}_J^{l,m}\boldsymbol{j}$ + $\mathbf{F}_K^{l,m}\boldsymbol{k}$.




To evaluate importance, $\mathit{I_\mathbf{F}^{l,m}}$ of $m^{th}$ quaternion filter in $l^{th}$ layer, we measure one of the following:

(a) In $\ell_1$-norm based pruning method \cite{li2017pruning},
the importance score of $m^{th}$ quaternion filter  is computed as:

\begin{equation}
\mathit{I}_\mathbf{F}^{l,m} = \lVert \mathbf{F}_R^{l,m} \rVert_1 + \lVert \mathbf{F}_I^{l,m} \rVert_1 + \lVert \mathbf{F}_J^{l,m} \rVert_1 + \lVert \mathbf{F}_K^{l,m} \rVert_1.
\end{equation}

(b) In Geometric-median (GM) based pruning method \cite{he2019filter}, 
we hypothesize that 
filters exhibiting minimal deviation from the geometric median are considered to encode redundant information
and may be pruned away.

The geometric-median \cite{he2019filter} is computed independently for each component for filters in QCNN. The geometric median of the \( o^{\text{th}} \) component, where \( o \in \{R, I, J, K\} \), is defined as:

\begin{equation}
\mathbf{F}_{o}^{\text{GM}} \in \underset{\mathbf{F}_{o}}{\text{argmin}} \sum_{j=1}^{M} \lVert \mathbf{F}_o - \mathbf{F}_o^{j} \rVert_2,
\label{equ:GM}
\end{equation}

\noindent where \( M \) denotes the total number of quaternion filters in the layer, and \( \mathbf{F}_{o}^{j} \) represents the \( o^{\text{th}} \) component of the \( j^{\text{th}} \) filter.

The importance of the \( m^{\text{th}} \) quaternion filter in the \( l^{\text{th}} \) layer is then computed as the sum of the \( \ell_1 \)-distances between its components and the corresponding component-wise geometric medians:

\begin{align}
\mathit{I}_\mathbf{F}^{l,m} = & \, \lVert \mathbf{F}_R^{l,m} - \mathbf{F}_R^{\text{GM}} \rVert_1 + \lVert \mathbf{F}_I^{l,m} - \mathbf{F}_I^{\text{GM}} \rVert_1 \nonumber \\
                    & + \lVert \mathbf{F}_J^{l,m} - \mathbf{F}_J^{\text{GM}} \rVert_1 + \lVert \mathbf{F}_K^{l,m} - \mathbf{F}_K^{\text{GM}} \rVert_1.
\end{align}





(c) In Operator-norm (OP) based pruning method \cite{singh_transaction}, we estimate the importance of each quaternion filter based on 
the largest singular value of a matrix derived from the filter weights.  

 
The overall importance score of the \( m^{\text{th}} \) filter in the \( l^{\text{th}} \) layer is computed as the sum of the operator norm computed across all four components:


\begin{align}
\mathit{I}_\mathbf{F}^{l,m} = \sum_{o \in \{R, I, J, K\}} \sup_{\mathbf{q} \ne 0} \frac{\|\mathbf{F}_{o}^{l,m}(\mathbf{q})\|}{\|\mathbf{q}\|} = \sum_{o \in \{R, I, J, K\}} \sigma_{1}^o,
\end{align}

\noindent where $\mathbf{q}$ denotes the quaternion input, $\sigma_{1}^{o}$ denotes the largest singular value obtained by SVD of the $o^{\text{th}}$ component $\mathbf{F}_{o}^{l,m}$, with $o \in \{R, I, J, K\}$.

\begin{algorithm}[t]
\caption{Layer-wise Quaternion Filter Pruning.}
\label{alg: filter pruning}
\KwIn{Pretrained model $\mathcal{M}$, pruning ratio $p$,  data $\mathcal{D}$}
\KwOut{Pruned and fine-tuned model $\mathcal{M}'$}

\ForEach{layer $l$ in convolutional layers of $\mathcal{M}$}{
    Compute filter importance using $\ell_l$-norm, geometric-median, or operator-norm pruning method,\\
    Prune $p$ least important filters in $l^{\text{th}}$layer.\
}
Obtain pruned model $\mathcal{M}'$,\

Fine-tune $\mathcal{M}'$ on $\mathcal{D}$ to recover performance.\

\Return{$\mathcal{M}'$}

\end{algorithm}

\section{Experimental setup}

\subsection{Datasets}
We evaluate  our approach on AudioSet (audio tagging task) \cite{gemmeke2017audio},
GTZAN (music genre classification) \cite{tzanetakis2002musical}, ESC-50 (environmental sound classification) \cite{piczak2015esc} and RAVDESS (speech emotion recognition) \cite{livingstone2012ravdess}.

\textbf{1) AudioSet:} AudioSet \cite{gemmeke2017audio}  is a large-scale audio dataset that contains over 2 million 10-second-long audio clips from 527 audio event classes.  Each 10-second audio clip can have multiple labels corresponding to different sound events occurring simultaneously. The dataset contains an unbalanced training set (approximately 2 million clips), and an evaluation set (20,383 clips).



\textbf{2) GTZAN:} GTZAN \cite{tzanetakis2002musical} is a music genre collection dataset  that consists of 1000 audio files, each of which is 30 seconds in duration. The dataset contains 10 classes representing 10 music genres: blues, classical, country, disco, hip-hop, jazz, metal, pop, reggae, and rock.


\textbf{3) ESC-50:} The ESC-50 dataset \cite{piczak2015esc} is a widely used benchmark for environmental sound classification (ESC). The dataset consists of 2,000 labelled audio clips, each 5 seconds long, recorded at 44.1 kHz, 16-bit WAV format. The sounds are categorized into 50 distinct classes that cover a diverse range of environmental, human, animal, and urban sounds.


\textbf{4) RAVDESS:} The Ryerson Audio-Visual Database of Emotional Speech and Song (RAVDESS) \cite{livingstone2012ravdess} is a widely used benchmark dataset for speech emotion recognition (SER). The dataset consists of 1,440 speech recordings, with an average duration of four seconds, from 24 professional actors (12 male, 12 female). Each actor expresses eight emotions: neutral, happy, sad, angry, fearful, disgusted, surprised, and calm. 


\subsection{Audio representation into multi-channels or quaternion}
\label{sec: audio rep into quaternion}



Inspired by Parcollet et al.~\cite{parcollet2018quaternion}, we represent each time--frequency bin of the log-mel spectrogram as a \textit{quaternion-valued signal} \( Q(f, t) \), where \( f \) and \( t \) denote the frequency bin and time frame, respectively. The quaternion encodes the mel-band energy \( \psi(f, t) \) along with its temporal dynamics, defined as:
\begin{equation}
Q(f,t) = \psi(f,t) 
        + \frac{\partial \psi(f,t)}{\partial t} \, \boldsymbol{i} 
        + \frac{\partial^2 \psi(f,t)}{\partial t^2} \, \boldsymbol{j} 
        + \frac{\partial^3 \psi(f,t)}{\partial t^3} \, \boldsymbol{k}.
\end{equation}
This formulation  represents multiple views of frequency $f$ at time frame $t$, consisting of  the static energy $\psi(f,t)$ and its local temporal variations into a single quaternion representation.

\subsection{Evaluation metrics and setup}


For AudioSet, we use the mean Average Precision (mAP) as the performance metric. The mAP is obtained by averaging the average precision over all classes and is defined as

\begin{equation}
\text{mAP} = \frac{1}{G} \sum_{g=1}^{G} \text{AP}_g,
\end{equation}

\noindent where $G$ denotes the total number of classes and $\text{AP}_g$ is the average precision for class $g$.
The models are trained for $10^{6}$ iterations with mixup data augmentation technique \cite{zhang2018mixup} using the unbalanced training set and evaluated on the evaluation set. Each experiment is repeated three times, and the reported results correspond to the mean across these runs.   

For other datasets (GTZAN, ESC-50, RAVDESS), we evaluate performance using classification accuracy, which is defined as the proportion of correctly classified examples, averaged across $\lambda$ folds:

\begin{equation}
\text{Accuracy} = \frac{1}{\lambda} \sum_{i=1}^{\lambda} 
\frac{\text{Correctly classified examples in fold } i}{\text{Total examples in fold } i}.
\label{eq:accuracy}
\end{equation}


We employ $\lambda$-fold cross-validation, where $\lambda=10$ for GTZAN, $\lambda=5$ for ESC-50 and $\lambda=4$ for RAVDESS. Each fold is trained for 50k iterations on the respective training split (90\%, 80\%, and 75\%) with mixup data augmentation technique \cite{zhang2018mixup} and evaluated on the remaining portion of the data. The final accuracy is reported as the mean across all $\lambda$ folds.

To evaluate model efficiency, we report the total number of learned parameters (\#Parameters), the number of multiply-accumulate operations (MACs), the energy consumption, and the estimated carbon emissions (CO$_2$ equivalent). For QCNNs, the total number of learned parameters is equal to the four times the number of quaternion parameters.
We compute energy and carbon emissions using the \textit{Machine Learning Impact calculator}~\cite{lacoste2019quantifying}, which estimates environmental impact based on hardware usage, runtime, and geographical energy intensity.

\subsection{Pre-trained QCNNs for pruning and knowledge distillation} 





We use two conventional convolutional architectures \cite{kong2020panns} as baselines: a feedforward CNN14 and a residual ResNet38. More details on CNN14 and residual ResNet38 is included in Section II of the \textcolor{black}{supplemental} material.
Using these baseline architectures, we design their quaternion equivalent, quaternion CNN14 (QCNN14) and quaternion ResNet38 (QResNet38). Subsequently, we train these quaternion models  on the multi-channel or quaternion representation of the audio dataset taken from the unbalanced training set of AudioSet. The trained quaternion models thus obtained are used  as original or baseline networks, and are either used as an unpruned network for pruning or as a teacher model in KD.




\subsection{Performance comparison}
\label{sec: performance compariosn}

\subsubsection{Comparison between  knowledge distillation and pruning}
To compare the performance of compressed QCNNs obtained through KD and pruning, we evaluate the following scenarios:

\begin{itemize}

        \item \textbf{Pruning with fine-tuning (Prune\_FT):}
        We apply data-independent filter pruning methods  followed by fine-tuning the pruned model as described in Section \ref{sec: filter pruning} and shown in Figure~\ref{fig: overall framework}(b) . The \(p \in \{0.25, 0.50, 0.75\}\) least important filters are removed to obtain the pruned QCNN. More details on layer-wise pruning for QCNN14 and QResNet38 is provided in  Section III of the \textcolor{black}{supplemental} material.
        
    \item \textbf{Knowledge Distillation (KD):} The compressed model is obtained using the KD pipeline shown in Figure~\ref{fig: overall framework}(a). We use the pre-trained QCNN14 or QResNet38 model, trained on the AudioSet dataset, as the teacher network. The student model architecture is designed to be the same as the architecture of the pruned QCNN obtained after pruning (\(p \in \{0.25, 0.50, 0.75\}\)) filters from the original QCNN. For KD experiments, we keep $\alpha~=$~0.5 and $T~=$~2 in Equation \ref{equ: kd_total}.
    
    

\end{itemize}

\subsubsection{Comparison with existing models}
\label{sec: other methods}
We compare the performance of compressed quaternion models with existing models. These include Transformer-based models such as AST \cite{gong2021ast}, generMERT \cite{lin2025learnable} and AST-Fusion \cite{niizumi2022composing}, and conventional CNNs such as CNN14 \cite{kong2020panns}, MobileNet models \cite{kong2020panns, schmid2023efficient, gong2021psla} and residual networks such as Efficient Residual Audio Neural Networks (ERANNs)~\cite{verbitskiy2022eranns}, and ResNet54\cite{kong2020panns}. More detail on these methods is included in Section IV of the \textcolor{black}{supplemental} material.


\section{Results and Analysis: Audio Tagging}
\label{sec: results}

\begin{table}[t]
\centering
\caption{Mean average precision (mAP) for QCNN14 and QResNet38, when student or pruned model is obtained at different pruning ratios in KD or Prune\_FT method.}
\setlength{\tabcolsep}{6pt} 
\renewcommand{\arraystretch}{1.15} 
\resizebox{0.475\textwidth}{!}{%
\begin{tabular}{@{}lccccc@{}}
\toprule
\multirow{2}{*}{\textbf{Model}} & \multirow{2}{*}{\textbf{Pruning ratio ($\bm{p}$)}} & \multirow{2}{*}{\textbf{KD}} & \multicolumn{3}{c}{\textbf{Prune\_FT}} \\ 
\cmidrule(lr){4-6}
 & & & $\bm{\ell}_1$ & GM & OP \\
\midrule
QCNN14   & Baseline (No KD/Pruning) & \multicolumn{4}{c}{0.428} \\
\cmidrule(lr){3-6}
         & 0.25 & 0.412 & 0.435 & 0.435 & 0.436 \\
         & 0.50 & 0.411 & 0.429 & 0.428 & 0.423 \\
         & 0.75 & 0.397 & 0.395 & 0.397 & 0.395 \\
\midrule
QResNet38 & Baseline (No KD/Pruning)  & \multicolumn{4}{c}{0.426} \\
\cmidrule(lr){3-6}
          & 0.25 & 0.406 & 0.422 & 0.422 & 0.420 \\
          & 0.50 & 0.403 & 0.419 & 0.417 & 0.421 \\
          & 0.75 & 0.402 & 0.410 & 0.410 & 0.409 \\
\bottomrule
\end{tabular}}
\textit{The standard deviation across all trials is \textcolor{black}{$<10^{-3}$}.}
\label{tab:combined_maps}
\end{table}

The mean average precision (mAP) obtained for the unpruned QCNN14 and the unpruned QResNet38 on AudioSet is 0.428 and 0.426 respectively. 
Table \ref{tab:combined_maps} compares the mAP obtained using KD and Prune\_FT methods for QCNN14 and QResNet38 models on AudioSet, when student or pruned model is obtained at different pruning ratios. We find that both models show reduced mAP as pruning increases, suggesting a trade-off between compression and performance.  The performance obtained using pruning methods is similar or better than KD method at various pruning ratios. Among the various pruning methods, we find that all the pruning methods perform similarly to each other. In \cite{singh_transaction}, we showed that operator-norm pruning method has better performance  compared to other pruning methods when pruning conventional CNNs. Therefore, we use operator-norm based pruning method hereafter for our Prune\_FT method.

\begin{figure}[t]
    \centering
    \includegraphics[scale = 0.35]{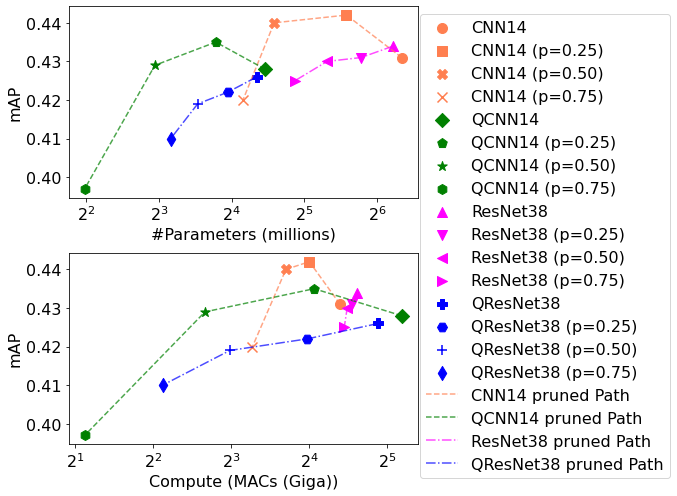}
    \caption{Comparison of mAP across conventional and quaternion, unpruned and pruned CNNs.}
    \label{fig: compre pruning cnn and qcnn}
\end{figure}

\subsection{Pruning quaternion versus conventional CNNs}

\begin{figure*}[ht]
    \centering
    \includegraphics[width =0.797\textwidth]{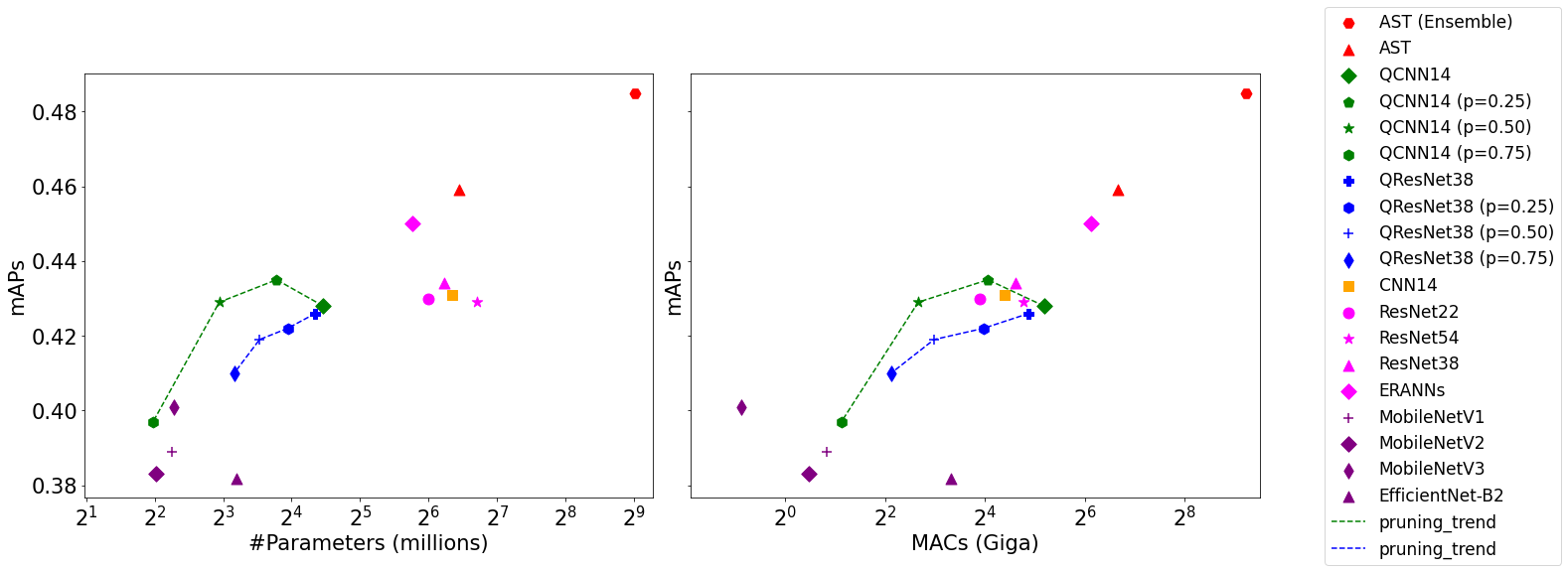}
    \caption{Pruned quaternion CNNs performance comparison with existing conventional CNNs and Transformer based models for Audio tagging.}
    \label{fig: compariosn with other methods}
\end{figure*}


The effect of pruning quaternion and conventional CNNs with results shown in Figure \ref{fig: compre pruning cnn and qcnn}. The mAP obtained by  the pruned QCNN14 ($p$ = 0.50),  which has 7.7M parameters and \textcolor{black}{6.34G} MACs, is one percentage point higher than  the pruned CNN14 ($p$ = 0.75) with \textcolor{black}{17.8}M parameters and \textcolor{black}{9.7}G MACs. Moreover, the pruned QCNN14 ($p$ = 0.50) gives similar performance to the conventional (unpruned) CNN14 while using only one-tenth as many parameters and one-third as many MACs as used by the conventional CNN14. Although, the pruned CNN14 ($p$ = 0.50) improves by one percentage point over the  unpruned CNN14, it requires three times as many  parameters and  twice as many MACs as the pruned QCNN14 ($p$ = 0.50).  

Similarly, the pruned QResNet38 ($p$ = 0.50) achieves slightly better mAP than the pruned ResNet38 ($p$ = 0.75) while using significantly fewer parameters and MACs. Compared to the unpruned ResNet38, the pruned ResNet38 ($p$ = 0.25) retains a higher level of performance than the pruned QResNet38 ($p$ = 0.25). However, it requires approximately 3.5 times as many parameters and 1.5 times as many MACs as the pruned QCNN14 ($p$ = 0.25).

These results suggest that pruning applied to quaternion CNNs yields both parameter-efficient and compute-efficient  pruned models compared to those obtained  after pruning conventional CNNs.

\subsection{Comparison with existing models}

Next, we analyse the  trade-off between performance and  the number of parameters or MACs  across various existing  neural network architectures 
designed for audio tagging task described in Section \ref{sec: other methods}.

As shown in Figure \ref{fig: compariosn with other methods}, we find that large Transformer-based models, such as the Audio Spectrogram Transformer (AST) \cite{gong2021ast} and its ensemble version, achieve the highest mAP scores. However, Transformer models use a significantly higher number of parameters and MACs than the convolutional models. 

The Efficient Residual Audio Neural Networks (ERANNs) model \cite{verbitskiy2022eranns}, which is based on efficient residual networks, uses fewer parameters and MACs than Transformer models. The ERANNs model performs better than other convolutional models such as ResNet22, ResNet38, ResNet54, and CNN14, however with higher MACs than these models. 
Lightweight models, including MobileNet (V1, V2, V3) and EfficientNet-B2, use fewer parameters and have a fewer computational cost, but have lower performance than the larger models.

In contrast to the existing models shown in Figure  \ref{fig: compariosn with other methods}, we observe that  pruned quaternion models provide a balanced trade-off among the number of parameters, MACs, and performance. Although, the Transformer models outperform QCNN14 by 3-5 percentage points, the QCNN14 model pruned at $p$ = 0.50  has  at least 16 times fewer number of parameters and MACs than the that of Transformer models \cite{gong2021ast}, while the QCNN14 model at $p$ = 0.50  still performing better than the MobileNet models. 

\section{Generalization analysis- transferring to other tasks} 
\label{sec: GTZAN MUSIC experimental section}






In this section, we evaluate pruned quaternion models on three additional audio classification tasks: GTZAN (music-genre), ESC-50 (environmental sound), and RAVDESS (speech emotion). We use pruned models obtained from the unpruned pre-trained baselines and, without any post-pruning fine-tuning. We train the pruned models on each dataset to analyse their generalization.


We compare our pruned quaternion models with  those existing CNNs and Transformer models which have publicly availability of results on the respective datasets in Table \ref{tab:generalization_performance_comparison}. For GTZAN, we compare our pruned quaternion models with genreMERT \cite{lin2025learnable}, CNN14 \cite{kong2020panns}, and AST-Fusion \cite{niizumi2022composing}; for ESC-50, we use CNN14 \cite{kong2020panns}, AST \cite{gong2021ast}, and ERANNs \cite{verbitskiy2022eranns} and for RAVDESS, we compare with CNN14 and ERANNs.

\begin{table}[t]
\centering
\caption{Performance Comparison across GTZAN, ESC-50 and RAVDESS datasets.}
\large
\resizebox{0.45\textwidth}{!}{%
\begin{tabular}{lccc}
\toprule
\multicolumn{4}{c}{\textbf{GTZAN, Music Genre Classification}} \\ 
\midrule
\textbf{Model} & \textbf{Accuracy (\%)} & \textbf{\#Parameters (millions)} & \textbf{MACs (Giga)}  \\ 
\midrule
AST-Fusion \cite{lin2025learnable,niizumi2022composing} & 90.0 & 89.57 &   $>$100.00\\ 
CNN14 \cite{kong2020panns}  & 91.5 & 80.00 &  ~~~~63.51 \\ 
genreMERT \cite{lin2025learnable} & 92.0 & 94.40 & $>$100.00 \\ 
Ours ~~(QCNN14, $p$ = 0.75)  & 91.0 & ~\textbf{3.94} &  ~~~~~\textbf{4.28} \\ 
Ours ~~(QCNN14, $p$ = 0.25)  & 93.3  & 13.76 & ~~~~18.88 \\ 
Ours (QResNet38, $p$ = 0.25)  & 94.3  &  15.38 &  ~~~~17.79\\
Ours (QResNet38, $p$ = 0.75)  & \textbf{94.9}  & ~\textbf{8.95} &  ~~~~~\textbf{6.45}\\
\midrule
\multicolumn{4}{c}{\textbf{ESC-50, Environmental Sound Classification}} \\ 
\midrule
\textbf{Model} & \textbf{Accuracy (\%)} & \textbf{\#Parameters (millions)} & \textbf{MACs (Giga)} \\ 
\midrule
CNN14 \cite{kong2020panns}  & 94.70 & 79.80 &~~~~10.52\\ 
AST \cite{gong2021ast}  & 95.70 & 87.00 & ~~~~59.23\\ 
ERANNs \cite{verbitskiy2022eranns} & 96.10 & 37.90 & ~~~~35.47 \\
Ours ~~(QCNN14, $p$ = 0.75)   & 94.60  & ~\textbf{3.96}& ~~~~~\textbf{1.65}\\ 
Ours ~~(QCNN14, $p$ = 0.25)   & 95.70 & 13.78 &  ~~~~16.17  \\ 
Ours (QResNet38, $p$ = 0.75)  & 97.39  &~8.95 &~~~~~3.82 \\
Ours (QResNet38, $p$ = 0.25)  & \textbf{97.60}  &  \textbf{15.38} &~~~~\textbf{15.16}\\
\midrule
\multicolumn{4}{c}{\textbf{RAVDESS, Speech Emotion Recognition}} \\ 
\midrule
\textbf{Model} & \textbf{Accuracy (\%)} & \textbf{\#Parameters (millions)} & \textbf{MACs (Giga)} \\ 
\midrule
CNN14 \cite{kong2020panns}  & 72.10 & 79.00 &  ~~~~~8.43\\  
ERANNs \cite{verbitskiy2022eranns} & 74.80 & 24.00 & ~~~~27.85\\
Ours (QResNet38, $p$ = 0.25)  & 71.73  &  15.38 & ~~~~15.06\\
Ours ~~(QCNN14, $p$ = 0.75)  & 72.77 & ~\textbf{3.94} & ~~~~~\textbf{1.55}\\ 
Ours (QResNet38, $p$ = 0.75)  & 73.61  & ~8.95 & ~~~~~3.72\\
Ours~~(QCNN14, $p$ = 0.25)  & \textbf{74.16}  & \textbf{13.75} & ~~~~\textbf{16.07}\\ 
\bottomrule
\end{tabular}%
}
\label{tab:generalization_performance_comparison}
\end{table}



On the GTZAN dataset, the QResNet38 ($p$ = 0.75) achieves the highest accuracy of 94.9\% using only 8.95M parameters and 6G MACs, while other models like CNN14 and genreMERT require over 80M parameters and over 60G MACs. On the ESC-50 dataset, the QResNet38 ($p$ = 0.25) gives 97.60\% accuracy outperforming larger models like AST  and ERANNs while using significantly reduced parameters and MACs. For speech emotion recognition on RAVDESS, the pruned  quaternion models also achieve competitive or superior accuracy compared to CNN14 and ERANNs, while significantly reducing parameter count and MACs. More results on confusion matrices, convergence plots for GTZAN, ESC-50 and RAVDESS using pruned quaternion models are included in  Section V (Figure S4-S9) of the supplemental material.

These results suggests  that our pruned quaternion models generalize well across different datasets. Pruned quaternion models can be  lightweight alternative to conventional CNN-based and Transformer-based architectures across multiple audio classification tasks.

\section{Environmental impact of compressed \& uncompressed models}

Table \ref{tab:model_comparison} compares the performance, inference time, energy consumption, and carbon emissions for various models evaluated 
on NVIDIA GPU (GeForce RTX 2070 with Max-Q Design)
using 1000 audio samples of 30-seconds length. 

The conventional ResNet38 has an inference time of $22.24$s, an energy usage of $6.6 \times 10^{-4}$ kWh and a carbon emission $1.6 \times 10^{-4}$ kg, while its compressed quaternion variants generally show reduced inference time, energy and carbon emissions as compression increases. Similarly, the compressed QCNN14 has significant reductions in inference time, energy, and carbon emissions compared to that of conventional CNN14.


Overall, quaternion conversion and compression for ResNet38  improves the inference time by 18\% and reduces the environmental impact by 18.75\%  with a performance drop of less than 2.5 percentage points compared to that  of uncompressed network. For CNN14, quaternion conversion and compression  improves the inference time 55\% by and reduces the environmental impact by 53\%  with a performance drop of less than 4 percentage points compared to that  of conventional CNN14.

\begin{table}[t]
\centering
\caption{mean average precision, Inference time, energy consumption, and carbon emissions for various models.}
\large
\resizebox{0.45\textwidth}{!}{%
\begin{tabular}{@{\hskip 2pt}l
                S[table-format=1.3]
                S[table-format=2.2]
                S[table-format=1.2]
                S[table-format=1.2]
                @{\hskip 2pt}}
\toprule
\textbf{Model} & \textbf{mAP} & \textbf{Time (s)} & \textbf{Energy  ($\times 10^{-4}$ kWh)} & \textbf{CO$_2$eq ($\times 10^{-4}$ kg)} \\
\midrule
ResNet38                & 0.434 & 22.24 & 6.6 & 1.6 \\
QResNet38               & 0.426 & 28.31 & 8.5 & 2.0 \\
QResNet38 ($p$ = 0.25)  & 0.422 & 23.99 & 7.1 & 1.7 \\
QResNet38 ($p$ = 0.50)  & 0.421 & 19.70 & 6.1 & 1.5 \\
QResNet38 ($p$ = 0.75)  & 0.410 & 18.22 & 5.4 & 1.3 \\
\midrule
CNN14                   & 0.431 & 21.10 & 4.7 & 1.5 \\
QCNN14                  & 0.428 & 29.87 & 8.9 & 2.1 \\
QCNN14 ~($p$ = 0.25)     & 0.436 & 20.92 & 6.3 & 1.4 \\
QCNN14 ~($p$ = 0.50)     & 0.429 & 13.69 & 4.1 & 1.0 \\
QCNN14 ~($p$ = 0.75)     & 0.397 & 9.51  & 2.8 & 0.7 \\
\bottomrule
\end{tabular}%
}
\label{tab:model_comparison}
\end{table}


\section{Ablation study}

\subsection{Multi-channel conventional CNNs versus quaternion CNNs}

We perform an ablation study to analyse whether the advantage of using pruned QCNNs over conventional CNNs is due to the multi-channel (quaternion) representation of the audio signal alone, or it is due to the combination of both the quaternion model and the multi-channel (quaternion) representation. For this, we train conventional CNNs with multi-channel (quaternion) input as described in Section~\ref{sec: audio rep into quaternion} and compare its  performance with that of the quaternion CNNs in Table \ref{tab: multi channel cnn_qcnn_comparison}  for audio tagging task. 

In Table \ref{tab: multi channel cnn_qcnn_comparison}, we find that unpruned QCNN14 performs similarly to the multi-channel conventional CNN14, while using approximately one-fourth as many parameters but 1.7 times as many MACs as that of multi-channel conventional CNN14.  Applying filter pruning to  QCNN14  reduces the MACs by 20\%  compared to multi-channel conventional CNN14 while using  only one-sixth  as many  parameters as that of the multi-channel conventional CNN14 with similar performance.

Similarly, QResNet38 performs similar to that of multi-channel conventional ResNet38 model while using approximately one-fourth as many parameters, but 1.2 times as many MACs as that of multi-channel conventional ResNet38. Applying filter pruning on QResNet38 reduces both its MACs and number of parameters significantly compared to that of multi-channel conventional ResNet38 with some loss of performance. 

The results in Table \ref{tab: multi channel cnn_qcnn_comparison} suggest that quaternion CNNs give similar performance at reduced number of parameters compared to that of multi-channel conventional CNNs. Pruning QCNN further reduces its number of parameters and computational cost. Therefore, the combination of quaternion and pruning makes quaternion models more efficient in terms of parameters and MACs for multi-channel input processing, while maintaining  comparable performance to multi-channel conventional CNNs. In future, we would like to analyse how model compression affects multi-channel conventional CNN performance. 

\begin{table}[t]
\centering
\caption{Comparison of performance between multi-channel conventional and quaternion CNNs for audio tagging task.}
\label{tab: multi channel cnn_qcnn_comparison}
\Large
\resizebox{0.49\textwidth}{!}{%
\begin{tabular}{lccc}
\toprule
\textbf{Model} & \textbf{Performance (mAP)} & \textbf{\#Parameters (millions)}  & \textbf{MACs (Giga)}\\
\midrule
Multi-channel CNN14 & 0.432 & 80.00 & 21.26 \\
QCNN14 (unpruned)  & 0.428 & 22.00  & 36.33 \\
QCNN14 ($p$ = 0.25)  & 0.436 & 13.78  & 16.70 \\
\midrule
Multi-channel ResNet38 &  0.425 & 75.00 &  24.55\\
QResNet38 (unpruned) & 0.426 & 20.36 & 29.51\\
QResNet38 ($p$ = 0.50) & 0.421 & 11.57 & ~7.90\\
\bottomrule
\end{tabular}%
}
\vspace{2.6mm}
\small \textit{The standard deviation across all  trials is  $< 10^{-3}$}.
\end{table}

\begin{figure}[t]
    \includegraphics[width =0.475\textwidth]{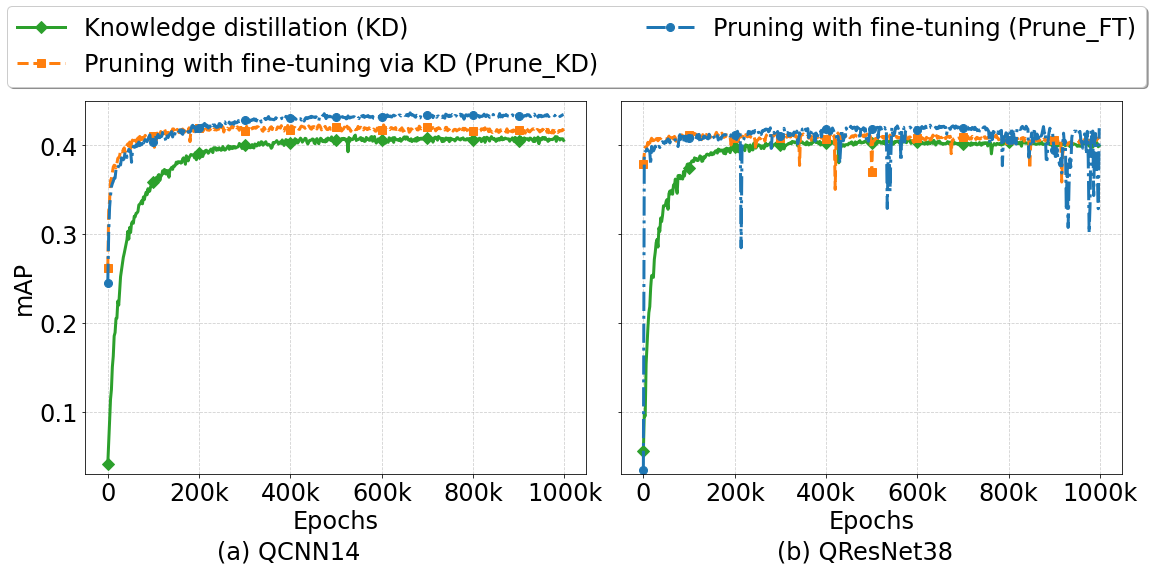}
    \caption{Convergence plots obtained across QCNN14 and QResNet38 for KD,   Prune\_FT  and Prune\_KD  for audio tagging task. The convergence plot is obtained by averaging the convergence plots across different pruning ratios using the operator norm-based pruning method.}
    \label{fig: pruning vs KD}
\end{figure} 

\subsection{Combining both pruning and knowledge distillation}

We analyse the performance of the pruned network when it is fine-tuned with KD. For this, similar to the Prune\_FT pipeline as mentioned in Section \ref{sec: performance compariosn}, we first apply filter pruning. Then,  we fine-tune the pruned model with knowledge distillation loss as given in Equation \ref{equ: kd_total}.

Figure \ref{fig: pruning vs KD} shows the mAP convergence obtained using KD, Prune\_FT, and  Prune\_KD (pruning with fine-tuning via KD) for  QCNN14 and QResNet38. We find that applying pruning and then fine-tuning with or without KD  converges faster while also giving better performance than applying  KD alone. The Prune\_FT consistently outperforms the Prune\_KD in terms of fine-tuning performance as shown in Figure~\ref{fig: pruning vs KD}. 

\begin{figure}[t]
    \centering
    \includegraphics[width=0.85\linewidth]{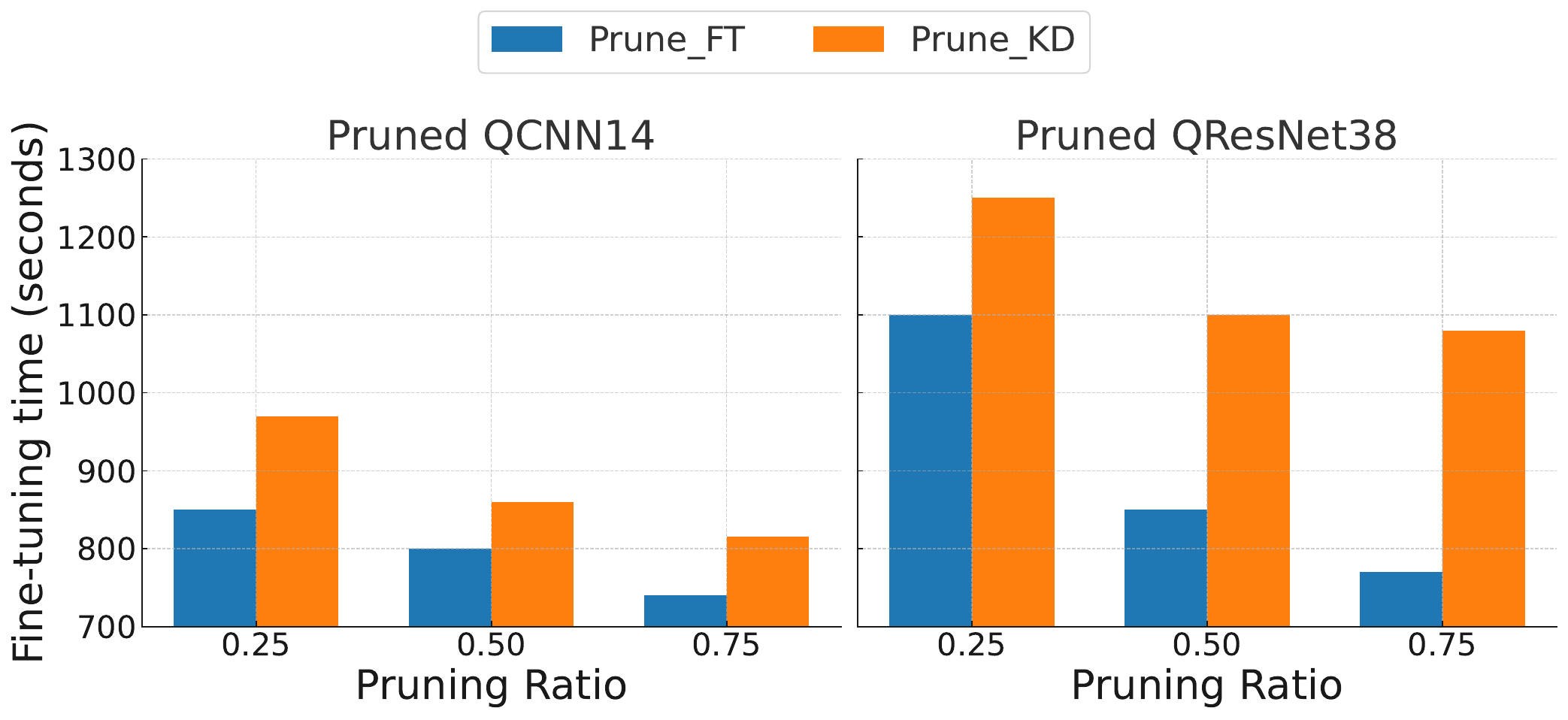}
    \caption{Comparison of average fine-tuning time per iteration for pruned QCNN14 and QResNet38 models across different pruning ratios (0.25, 0.50, 0.75) using Prune\_FT (pruning with fine-tuning) and Prune\_KD (pruning with fine-tuning via KD) methods for audio tagging task. The models are trained on NVIDIA GPU (GeForce RTX 3090).}
    \label{fig: fine-tuning time}
\end{figure}

In fine-tuning the pruned model, Prune\_KD approach requires an additional forward pass of the teacher model~\cite{xu2023computation}, leading to increased computational cost compared to the Prune\_FT, as shown in Figure~\ref{fig: fine-tuning time}. These results suggest that fine-tuning a pruned model without KD is more compute-efficient than fine-tuning with KD.


\section{Conclusion}

In this work, we compress quaternion convolutional neural networks (QCNNs) to reduce their computational complexity and memory requirements. We explore knowledge distillation and pruning methods to compress QCNNs. Pruning methods are found to be performing better than that of knowledge distillation. Further, we find that
fine-tuning a pruned model without knowledge distillation is more compute-efficient than fine-tuning with knowledge distillation.
Applying pruning to quaternion CNNs produce parameter-efficient and compute-efficient models when compared to pruning conventional CNNs.

On AudioSet, a pruned QCNN14  gives a good balance of parameters, computations and performance compared to large-scale Transformer and lightweight MobileNet models.
Pruned QCNNs perform similar to multi-channel conventional CNNs while using a reduced  parameter count and computational cost. Pruned QCNNs also generalize well across various audio classification tasks, outperforming existing methods for music genre classification and environmental sound classification, and performing similarly to  existing methods for speech emotion recognition while using a significantly reduced number of parameters and computational cost. Compression of QCNNs also reduce energy consumption and carbon emissions, contributing to the development of more sustainable AI models.




In future work, we would like to study the trade-off between performance and computation while using knowledge distillation  with large-scale teacher models, such as large Transformers or ensembles of Transformer models. We also would like to explore pruning in quaternion Transformer, multi-channel conventional CNNs and complex-valued networks.



\section{Acknowledgement}
This work was supported by Engineering and Physical Sciences Research Council (EPSRC) Grant EP/T019751/1 \enquote{AI for Sound (AI4S)}. For the purpose of open access, the authors have applied a Creative Commons Attribution (CC BY) licence to any Author Accepted Manuscript version arising. The authors would like to thank to Aryan Chaudhary for initial discussions.

\clearpage

\section*{Supplemental Material:  Compressing Quaternion Convolutional Neural Networks for Audio Classification}

\author{Arshdeep Singh, Vinayak Abrol, \IEEEmembership{Senior Member, IEEE} and Mark D. Plumbley \IEEEmembership{Fellow, IEEE}}

\maketitle

\setcounter{section}{0}
\renewcommand{\thesection}{\Roman{section}}
\setcounter{figure}{0}  
\renewcommand{\thefigure}{S\arabic{figure}}
\setcounter{table}{0}  
\renewcommand{\thetable}{S\arabic{table}}

\section{Hamilton product and convolutions in Conventional, Quaternion CNNs}

\label{sec: background: quaternion convolution}
The Hamilton product ($\otimes$) between two quaternions $\mathbf{m} = m_R + m_I\boldsymbol{i} + m_J\boldsymbol{j} + m_K\boldsymbol{k}, \mbox{ and}\quad \mathbf{p} = p_R + p_I\boldsymbol{i} + p_J\boldsymbol{j} + p_K\boldsymbol{k}$ is expressed as

\[
\mathbf{m} \otimes \mathbf{p} =
\begin{bmatrix}
m_R & -m_I & -m_J & -m_K \\
m_I &  m_R & -m_K &  m_J \\
m_J &  m_K &  m_R & -m_I \\
m_K & -m_J &  m_I &  m_R \\
\end{bmatrix}
\begin{bmatrix}
p_R \\
p_I \\
p_J \\
p_K \\
\end{bmatrix}.
\]

Quaternion CNNs extend the scalar operations of conventional convolution  to the quaternion convolution  by replacing scalar multiplications with the Hamilton product.

To illustrate the difference between convolution operation in conventional and quaternion CNNs, consider a 2D convolutional layer with \( C_{\text{in}} = 4 \) input channels and \( C_{\text{out}} = 4 \) output channels. In a conventional convolutional neural network (CNN), the input feature map is represented by a 3D tensor \( \mathbf{X} \in \mathbb{R}^{C_{\text{in}} \times H \times W} \), where \( H \) and \( W \) denote the height and width of the input. The  conventional convolutional kernel matrix is represented by \( \mathbf{W} \in \mathbb{R}^{C_{\text{out}} \times C_{\text{in}} \times k_h \times k_w} \), where \( k_h \) and \( k_w \) are the height and width of the convolutional filters. The output is denoted by \( \mathbf{Y} \in \mathbb{R}^{C_{\text{out}} \times H' \times W'} \), where \( H' \) and \( W' \) correspond to the spatial dimensions of the output after convolution.

In conventional convolution, each output channel is computed as a sum over convolutions with the input channels:

\[
\mathbf{Y}_{d,:,:} = \sum_{c=1}^{C_{\text{in}}} \mathbf{W}_{d,c,:,:} \mathbf{X}_{c,:,:}, \quad \text{for } d = 1, \dots, C_{\text{out}},
\]

\noindent which can be represented in matrix form as:

\begin{equation}
\begin{bmatrix}
\mathbf{Y}_1 \\
\mathbf{Y}_2 \\
\mathbf{Y}_3 \\
\mathbf{Y}_4
\end{bmatrix}
=
\begin{bmatrix}
\mathbf{W}_{1,1} & \mathbf{W}_{1,2} & \mathbf{W}_{1,3} & \mathbf{W}_{1,4} \\
\mathbf{W}_{2,1} & \mathbf{W}_{2,2} & \mathbf{W}_{2,3} & \mathbf{W}_{2,4} \\
\mathbf{W}_{3,1} & \mathbf{W}_{3,2} & \mathbf{W}_{3,3} & \mathbf{W}_{3,4} \\
\mathbf{W}_{4,1} & \mathbf{W}_{4,2} & \mathbf{W}_{4,3} & \mathbf{W}_{4,4}
\end{bmatrix}
\begin{bmatrix}
\mathbf{X}_1 \\
\mathbf{X}_2 \\
\mathbf{X}_3 \\
\mathbf{X}_4
\end{bmatrix}
\label{eq: real_conv_matrix}
\end{equation}

\noindent where  $\mathbf{X}_c \in \mathbb{R}^{H \times W}$, $\mathbf{Y}_d \in \mathbb{R}^{H' \times W'}$. Assuming \( k_h = 3 \) and \( k_w = 3 \),  each convolutional kernel will be $\mathbf{W}_{d,c} \in \mathbb{R}^{3 \times 3}$. There are 16 such independent kernels in Equation \ref{eq: real_conv_matrix},  the  number of learned parameters is calculated as $16 \times 3 \times 3 = 144$.

\medskip

In quaternion-valued convolution, the four input channels are grouped to form a single quaternion-valued input:

\[
\mathbf{Q} = \mathbf{X}_R + \mathbf{X}_I \boldsymbol{i} + \mathbf{X}_J \boldsymbol{j} + \mathbf{X}_K \boldsymbol{k}, \in \mathbb{H},
\]

\noindent where ($\mathbf{X}_R  , \mathbf{X}_i, \mathbf{X}_j, \mathbf{X}_k) \equiv (\mathbf{X}_1, \mathbf{X}_2, \mathbf{X}_3, \mathbf{X}_4)$. 
Quaternion kernel matrix is defined as:


\[
\mathbf{F} = \mathbf{F}_R + \mathbf{F}_I \boldsymbol{i} + \mathbf{F}_J \boldsymbol{j} + \mathbf{F}_K \boldsymbol{k},  \in \mathbb{R}^{\frac{C_{\text{out}}}{4} \times \frac{C_{\text{in}}}{4} \times k_h \times k_w},
\]

\noindent and the output is computed using the Hamilton product:

\[
\mathbf{Y} = \mathbf{F} \otimes \mathbf{Q}, \in \mathbb{H}.
\]

\noindent Expanding the Hamilton product gives the matrix form of quaternion convolution:

\begin{equation}
\begin{bmatrix}
\mathbf{Y}_R \\
\mathbf{Y}_I \\
\mathbf{Y}_J \\
\mathbf{Y}_K
\end{bmatrix}
=
\begin{bmatrix}
\mathbf{F}_R & -\mathbf{F}_I & -\mathbf{F}_J & -\mathbf{F}_K \\
\mathbf{F}_I &  \mathbf{F}_R & -\mathbf{F}_K &  \mathbf{F}_J \\
\mathbf{F}_J &  \mathbf{F}_K &  \mathbf{F}_R & -\mathbf{F}_I \\
\mathbf{F}_K & -\mathbf{F}_J &  \mathbf{F}_I &  \mathbf{F}_R
\end{bmatrix}
\begin{bmatrix}
\mathbf{X}_R \\
\mathbf{X}_I \\
\mathbf{X}_J \\
\mathbf{X}_K
\end{bmatrix}
\label{eq:quaternion_conv_matrix}
\end{equation}


\noindent $\mathbf{F}_R, \mathbf{F}_I, \mathbf{F}_J, \mathbf{F}_K$ are reused across all quaternion components to produce different components of quaternion output. Every output component in Equation \ref{eq:quaternion_conv_matrix} depends on all four input components and all four kernel components. Therefore, the Hamilton product naturally unifies the mixing of channels as part of its computation to learn correlations and interactions among all input channels simultaneously.

Unlike convolution in conventional CNNs as given in Equation \ref{eq: real_conv_matrix}, that requires a separate kernel for each input-output channel pair, quaternion convolution as given in Equation \ref{eq:quaternion_conv_matrix} leverages the structured sharing of convolutional filters to achieve a four-fold reduction in the number of learned parameters. In Equation \ref{eq:quaternion_conv_matrix}, there are four  independent convolutional kernels  $\mathbf{F}_R, \mathbf{F}_I, \mathbf{F}_J, \mathbf{F}_K$ each of size ($3 \times 3$), when  \( C_{\text{in}} = 4 \)  and \( C_{\text{out}} = 4 \). Therefore,  the  number of learned parameters for quaternion convolution would be $4 \times 3 \times 3 = 36$, four times fewer than that of conventional convolution as given in Equation \ref{eq: real_conv_matrix}

\section{Basleline conventional CNNs and training quaternion CNNs}
We use two standard convolutional architectures \cite{kong2020panns} as baselines: a feedforward CNN14 and a residual ResNet38. Both models take a log-mel spectrogram input of size \(1000 \times 64\), corresponding to a 10-second audio clip sampled at 32\,kHz with a window size of 1024 samples and a hop size of 320 samples. Starting from the baseline architecture, we perform quaternion model conversion as described in Algorithm I. A brief overview of the models is given below:

\textbf{CNN14:} A feedforward CNN architecture comprising 12 convolutional layers, CNN14 has approximately 81 million parameters and requires 21G multiply-accumulate operations (MACs)\footnote{\href{https://pypi.org/project/thop/}{MACs computed using PyTorch THOP package.}}. On the AudioSet benchmark, CNN14 achieves a mean average precision (mAP) of 0.431~\cite{kong2020panns}.

\textbf{ResNet38:} A deeper residual convolutional network, ResNet38 comprises around 74 million parameters and 24G MACs. On AudioSet, ResNet38 yields an mAP of 0.434~\cite{kong2020panns}.

\begin{figure}[h]
    \centering
    \includegraphics[scale=0.45]{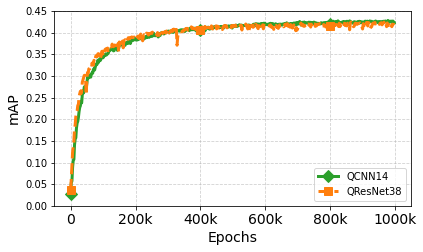}
    \caption{Convergence curve obtained during training of QCNN14 and QResNet38 on AudioSet.}
    \label{fig: training qcnns}
\end{figure}

\begin{figure}[ht]
    \centering
    \includegraphics[scale = 0.3]{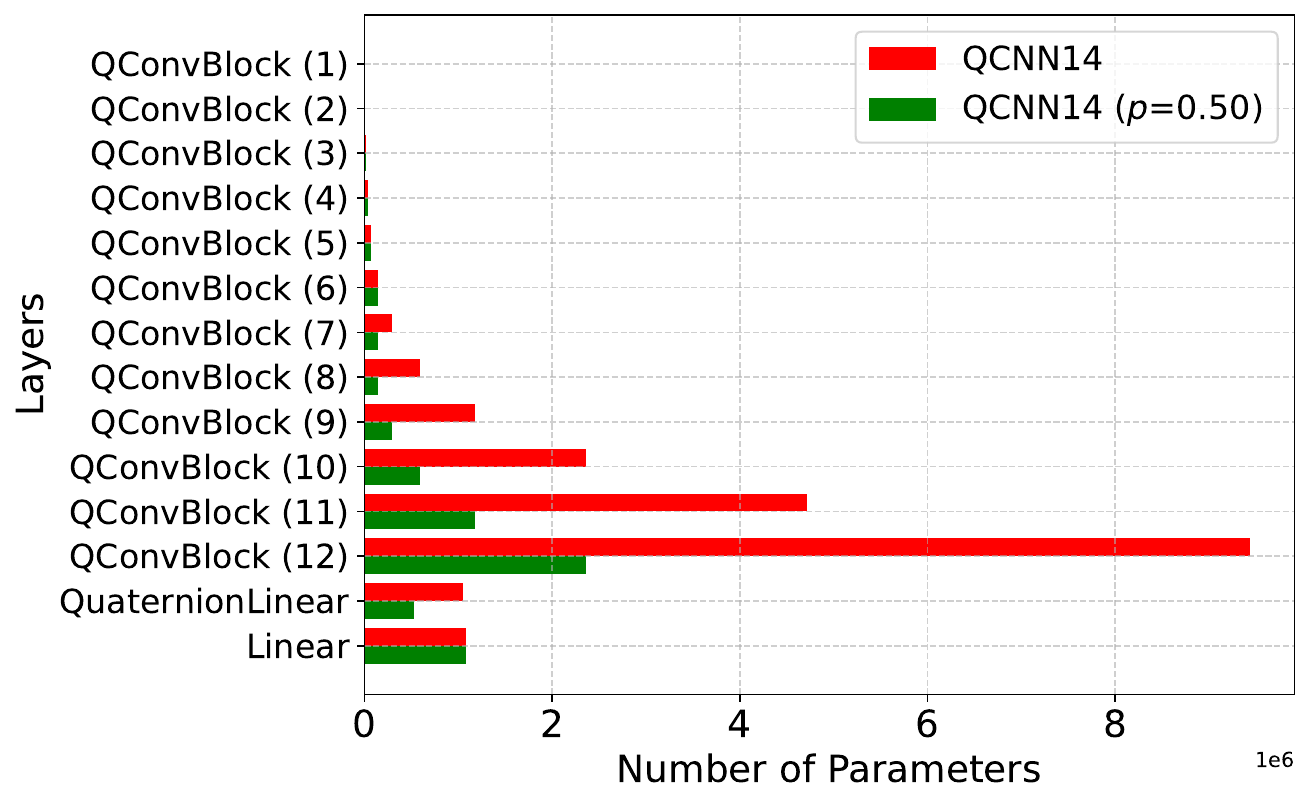}
    \caption{QCNN14 parameters across various intermediate layers before and after pruning.}
    \label{fig: QCNN14 parameters across layers}
\end{figure}

\begin{figure}[ht]
    \centering
    \includegraphics[scale = 0.3]{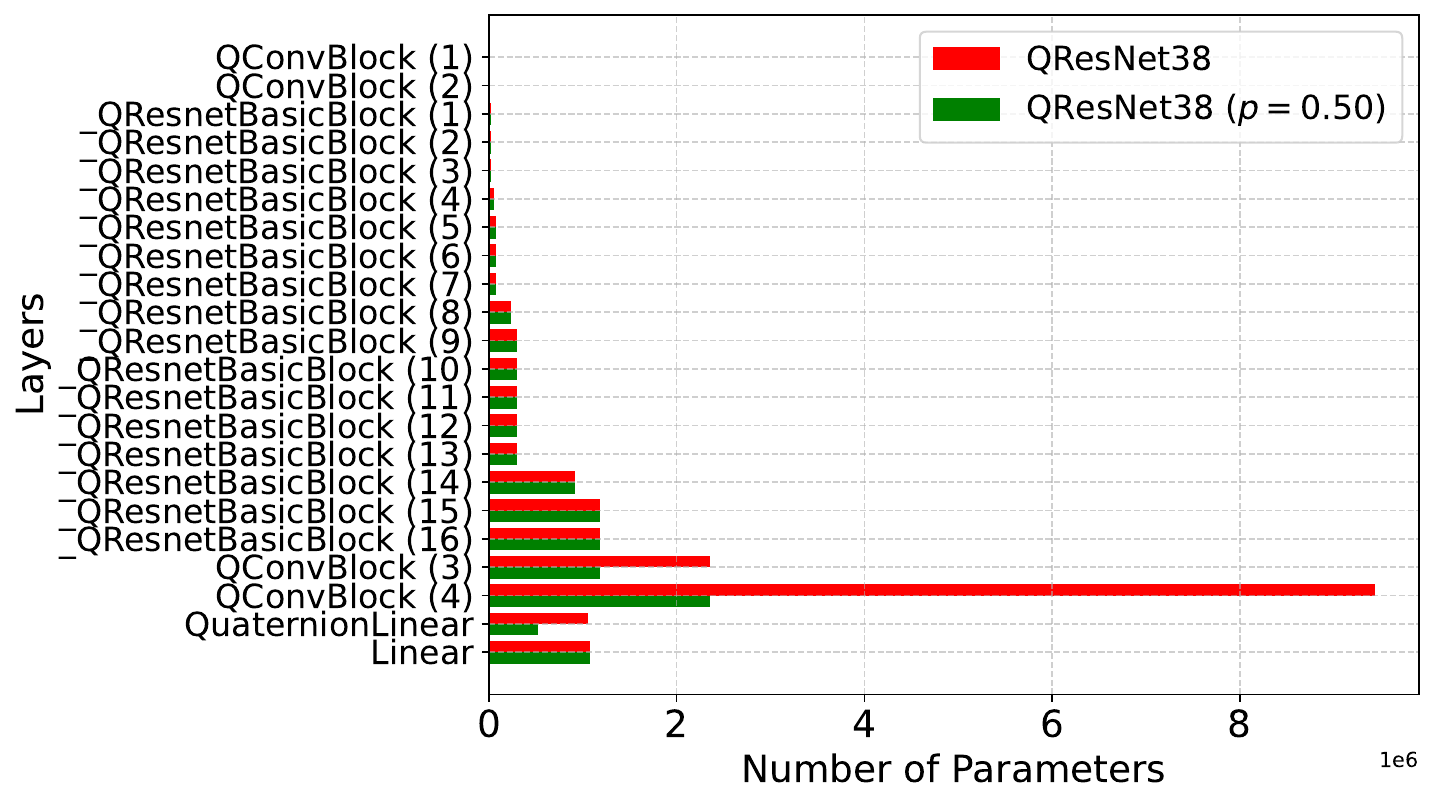}
    \caption{QResNet38 parameters across various intermediate layers before and after pruning.}
    \label{fig: QResNet parameters across layers}
\end{figure}

To train quaternion models, first, we convert the conventional CNNs as explained in the previous section to their quaternion equivalent. Subsequently, we perform training of the quaternion models using AudioSet dataset. The convergence curve obtained during the training process is shown in Figure \ref{fig: training qcnns}. 

\section{Layer-wise pruning of quaternion models}

After obtaining the trained quaternion models, we apply filter pruning across their various intermediate layers. For QCNN14, we prune intermediate layers from seventh quaternion convolutional layers onwards, as these layers contribute significantly to the total number of parameters as layer-wise analysis of QCNN14 shown in Figure \ref{fig: QCNN14 parameters across layers}. Similarly, for QResNet38, we prune last two Quaternion convolutional layers due to their significant contribution compared to other layers as shown in QResNet38 layer-wise parameters analysis in Figure \ref{fig: QResNet parameters across layers}. We did not apply pruning to residual blocks 
To eliminate quaternion filters, we apply the filter pruning strategies to measure the importance of layer-wise filters and then based on $p = \{0.25,0.50.0.75\}$, we eliminate top $p$\ unimportant quaternion filters. An illustration of parameters obtained after applying $p = 0.50$ is also shown in Figure \ref{fig: QCNN14 parameters across layers} and \ref{fig: QResNet parameters across layers}.

\section{A background on methods used for comparison}

A brief overview of existing methods used for comparison is  given below, 

\begin{itemize}
    \item \textbf{genreMERT}~\cite{lin2025learnable}: This is an attention-based architecture designed for acoustic music understanding. It leverages large-scale self-supervised pretraining using masked encoder representations from transformers (MERT) ~\cite{yizhi2023mert}—a transformer-based audio representation model—and introduces learnable counterfactual attention to improve robustness and interpretability in genre classification tasks.
    
    \item \textbf{AST}: The Audio Spectrogram Transformer (AST), proposed by~\cite{gong2021ast}, is a transformer-based architecture tailored for audio classification tasks. It adapts the Vision Transformer (ViT) to process audio by treating log-mel spectrograms as 2D inputs, dividing them into fixed-size time-frequency patches.
    
    \item \textbf{AST-Fusion}~\cite{niizumi2022composing}: An enhanced version of AST, this model fuses features from multiple intermediate transformer layers to improve performance across audio classification tasks. The fusion mechanism exploits the hierarchical nature of transformer representations for more effective learning.

    \item \textbf{Residual networks:} Residual networks are deep neural architectures that facilitate the training of very deep models by introducing shortcut connections to learn residual mappings, effectively addressing optimization challenges such as vanishing gradients. ResNet22, ResNet34, and ResNet54 \cite{kong2020panns} are deep convolutional neural networks with 22, 34, and 54 layers, respectively. Efficient Residual Audio Neural Networks (ERANNs)~\cite{verbitskiy2022eranns} are compact residual networks which uses residual connections, depthwise separable convolutions, and optimized filter configurations, providing a good balance between accuracy and computational efficiency compared to Transformer-based models \cite{gong2021ast} for audio classification task.
    

    \item \textbf{Lightweight models:} The MobileNet family, including V1~\cite{kong2020panns}, V2~\cite{kong2020panns}, and V3~\cite{schmid2023efficient}, is designed for efficient deep learning on mobile and embedded devices. MobileNetV1 introduces depthwise separable convolutions, significantly reducing computational cost compared to standard convolutions. MobileNetV2 builds on this by introducing inverted residual blocks and linear bottlenecks to further improve information flow and efficiency. MobileNetV3 leverages neural architecture search (NAS) and lightweight attention mechanisms, such as squeeze-and-excitation (SE) blocks, to optimize for both latency and accuracy. In contrast, EfficientNet-B2~\cite{gong2021psla} applies a compound scaling method to uniformly scale network depth, width, and resolution, achieving high accuracy with fewer parameters and lower computational complexity. Together, these models represent a class of architectures that balance model size, computational cost, and performance, making them suitable for resource-constrained environments
\end{itemize}

\newpage

\section{Experiments on GTZAN, ESC-50 \& RAVDESS}


\begin{figure}[h]
    \centering
    \begin{subfigure}{0.49\textwidth}
        \centering
        \includegraphics[scale=0.25]{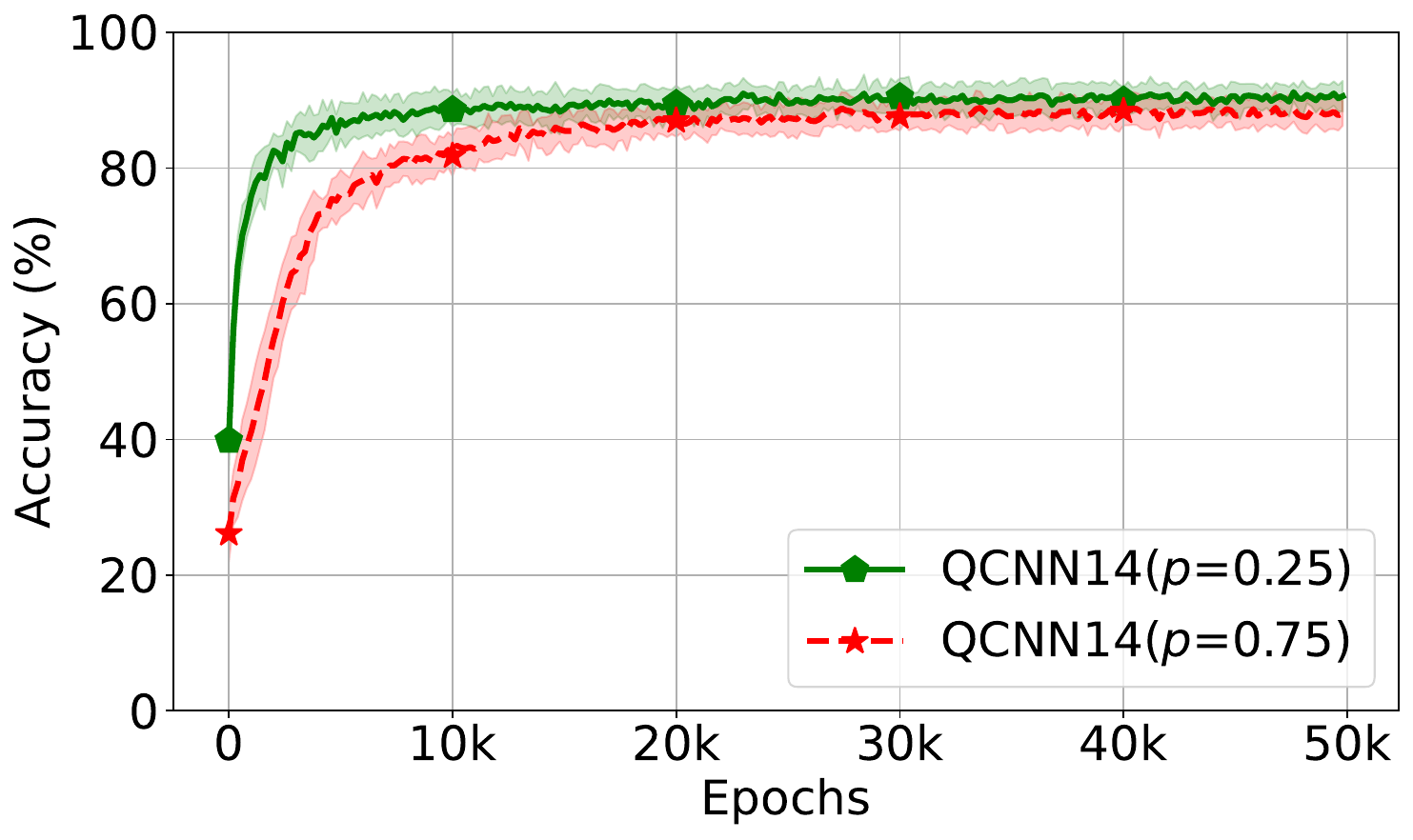}
        \caption{Convergence curve averaged across 10-folds for GTZAN.}
        \label{fig:qcnn25_GTZAN_convergence}
    \end{subfigure}
    \hfill
    \begin{subfigure}{0.49\textwidth}
        \centering
        \includegraphics[scale=0.25]{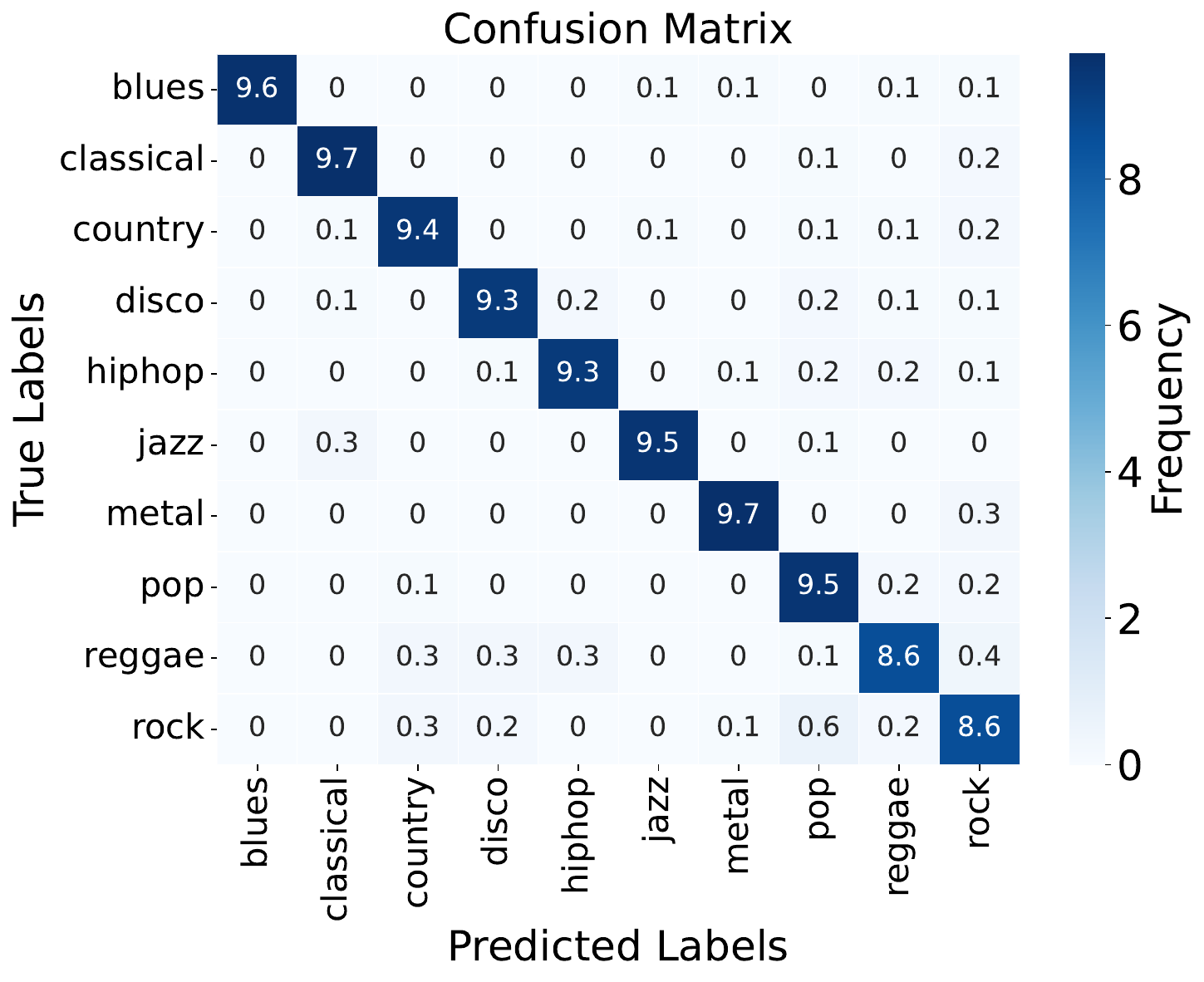}
        \caption{Averaged confusion matrix for QCNN14 ($p = 0.25$).}
        \label{fig:qcnn25_GTZAN_confusion}
    \end{subfigure}

    \hfill
    \begin{subfigure}{0.49\textwidth}
        \centering
        \includegraphics[scale=0.25]{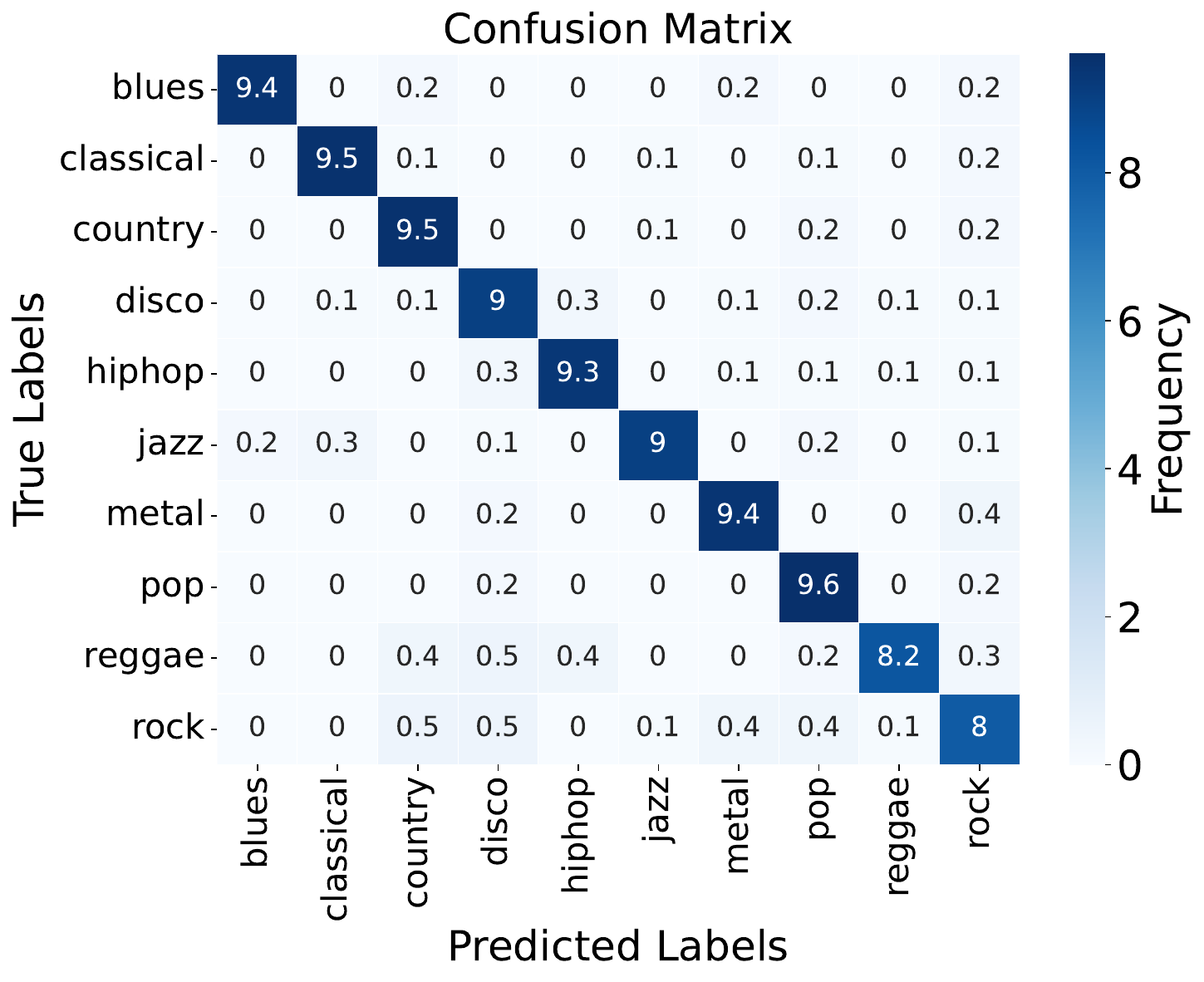}
        \caption{Averaged confusion matrix for QCNN14 ($p = 0.75$).}
        \label{fig:qcnn75_GTZAN_confusion}
    \end{subfigure}

    \caption{GTZAN dataset: (a) convergence curve   and (b, c)  average of confusion matrices (CM) over 10-folds. Each fold-wise CM is obtained from QCNN14, where it gives maximum accuracy.}
\end{figure}

\begin{figure}[h]
    \centering
    \begin{subfigure}{0.49\textwidth}
        \centering
        \includegraphics[scale=0.25]{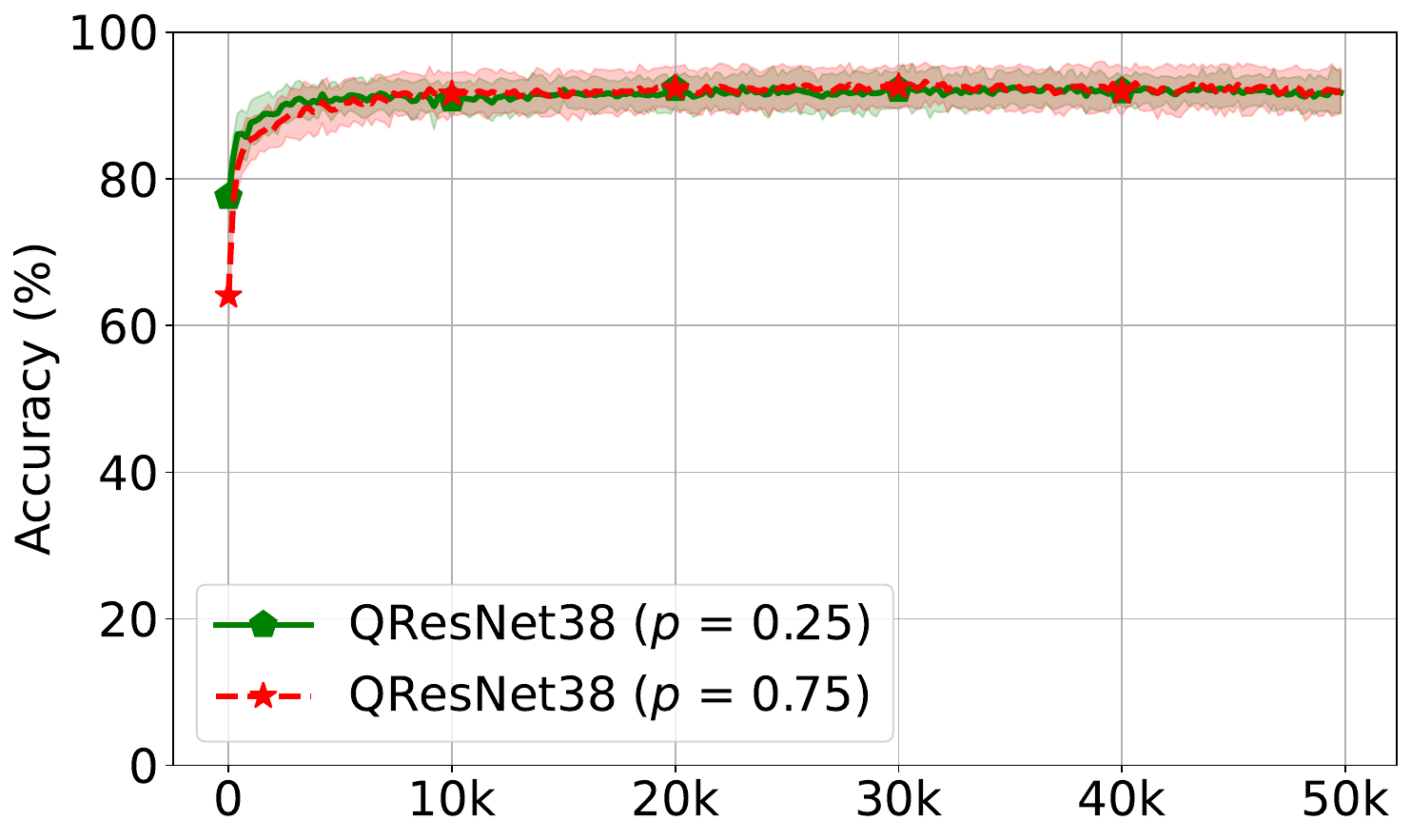}
        \caption{Convergence curve averaged across 10-folds for GTZAN.}
        \label{fig:qcnn25_GTZAN_convergence}
    \end{subfigure}
    \hfill
    \begin{subfigure}{0.49\textwidth}
        \centering
        \includegraphics[scale=0.25]{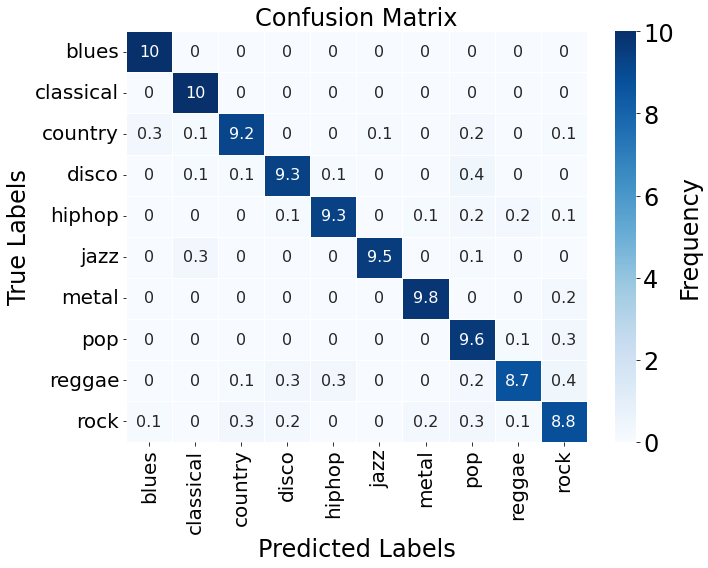}
        \caption{Averaged confusion matrix for QResNet38 ($p = 0.25$).}
        \label{fig:qcnn25_GTZAN_confusion}
    \end{subfigure}

    \hfill
    \begin{subfigure}{0.49\textwidth}
        \centering
        \includegraphics[scale=0.25]{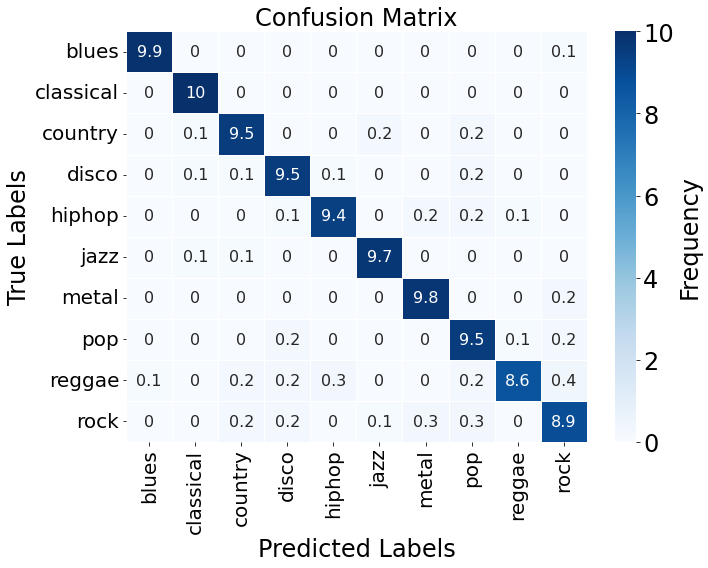}
        \caption{Averaged confusion matrix for QResNet38 ($p = 0.75$).}
        \label{fig:qcnn75_GTZAN_confusion}
    \end{subfigure}

    \caption{GTZAN dataset: (a) convergence curve   and (b, c)  average of confusion matrices (CM) over 10-folds. Each fold-wise CM is obtained from QResNet38, where it gives maximum accuracy}
\end{figure}


\begin{figure}[t]
    \centering
    \begin{subfigure}{0.49\textwidth}
        \centering
        \includegraphics[scale=0.25]{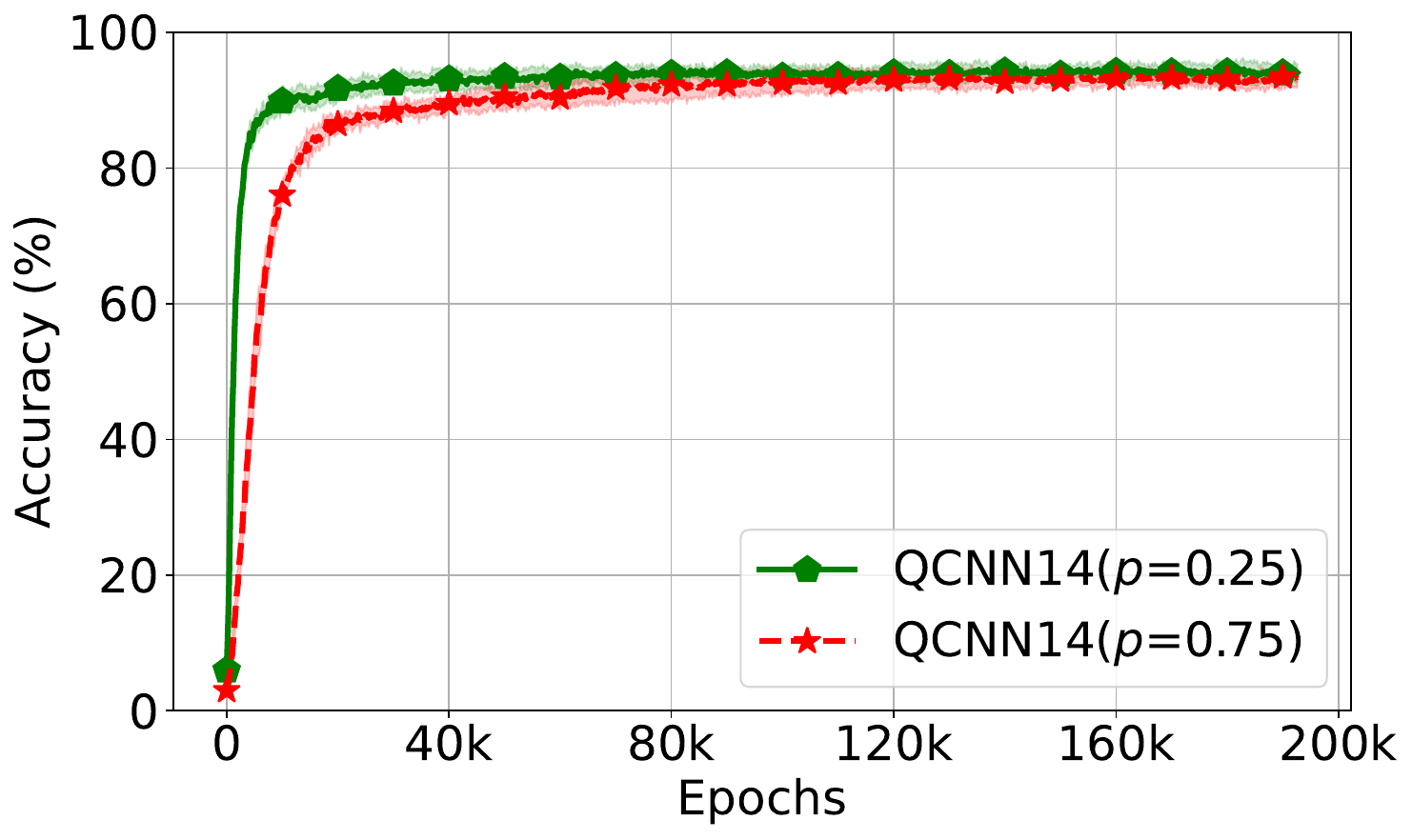}
        \caption{Convergence curve averaged across 5-folds for ESC-50.}
        \label{fig:ESC-50_qcnn25_convergence}
    \end{subfigure}
    \hfill
    \begin{subfigure}{0.49\textwidth}
        \centering
        \includegraphics[scale=0.15]{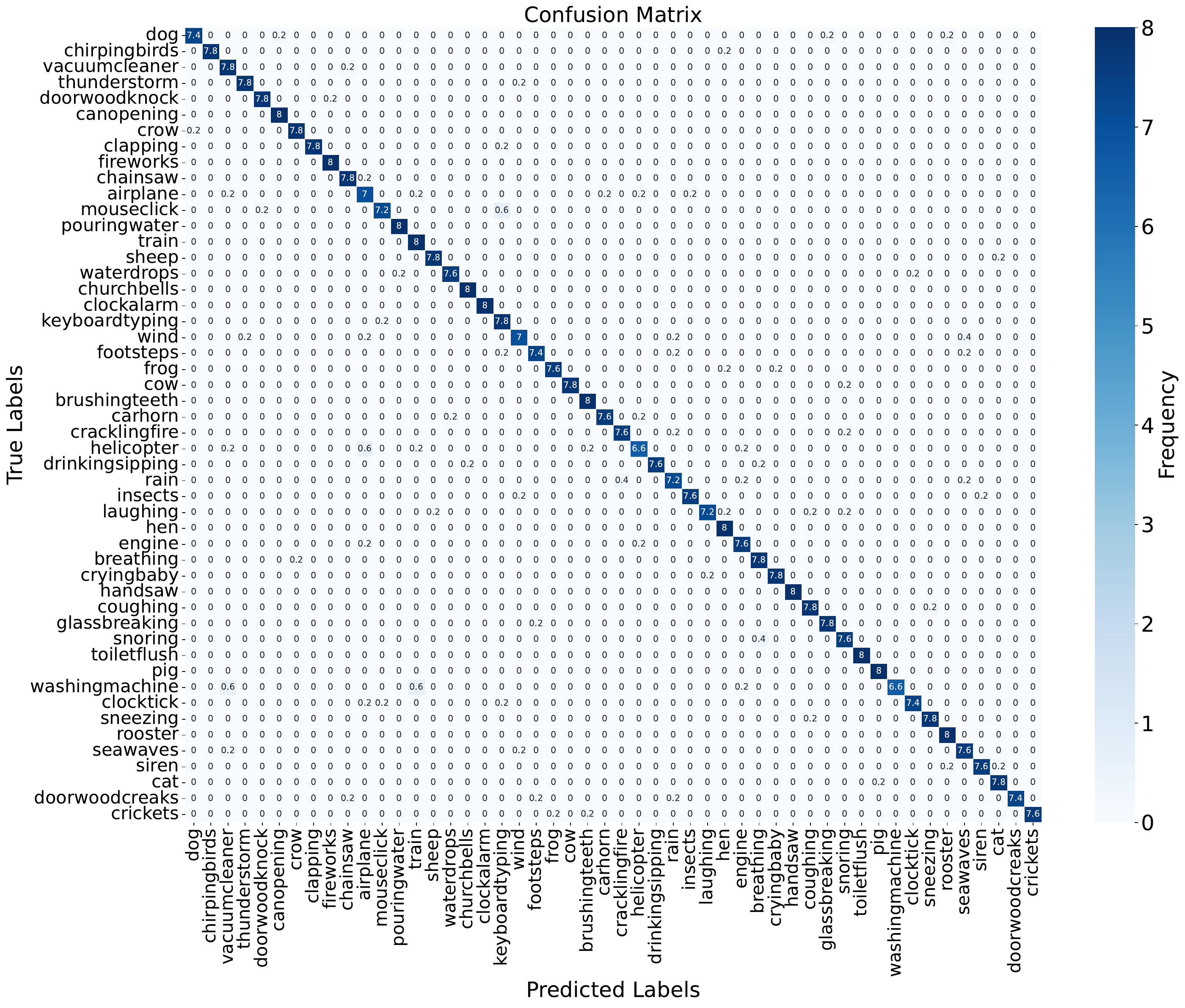}
        \caption{Averaged confusion matrix for QCNN14 ($p = 0.25$).)}
        \label{fig:ESC-50_qcnn25_confusion}
    \end{subfigure}

    \hfill
    \begin{subfigure}{0.48\textwidth}
        \centering
        \includegraphics[scale=0.15]{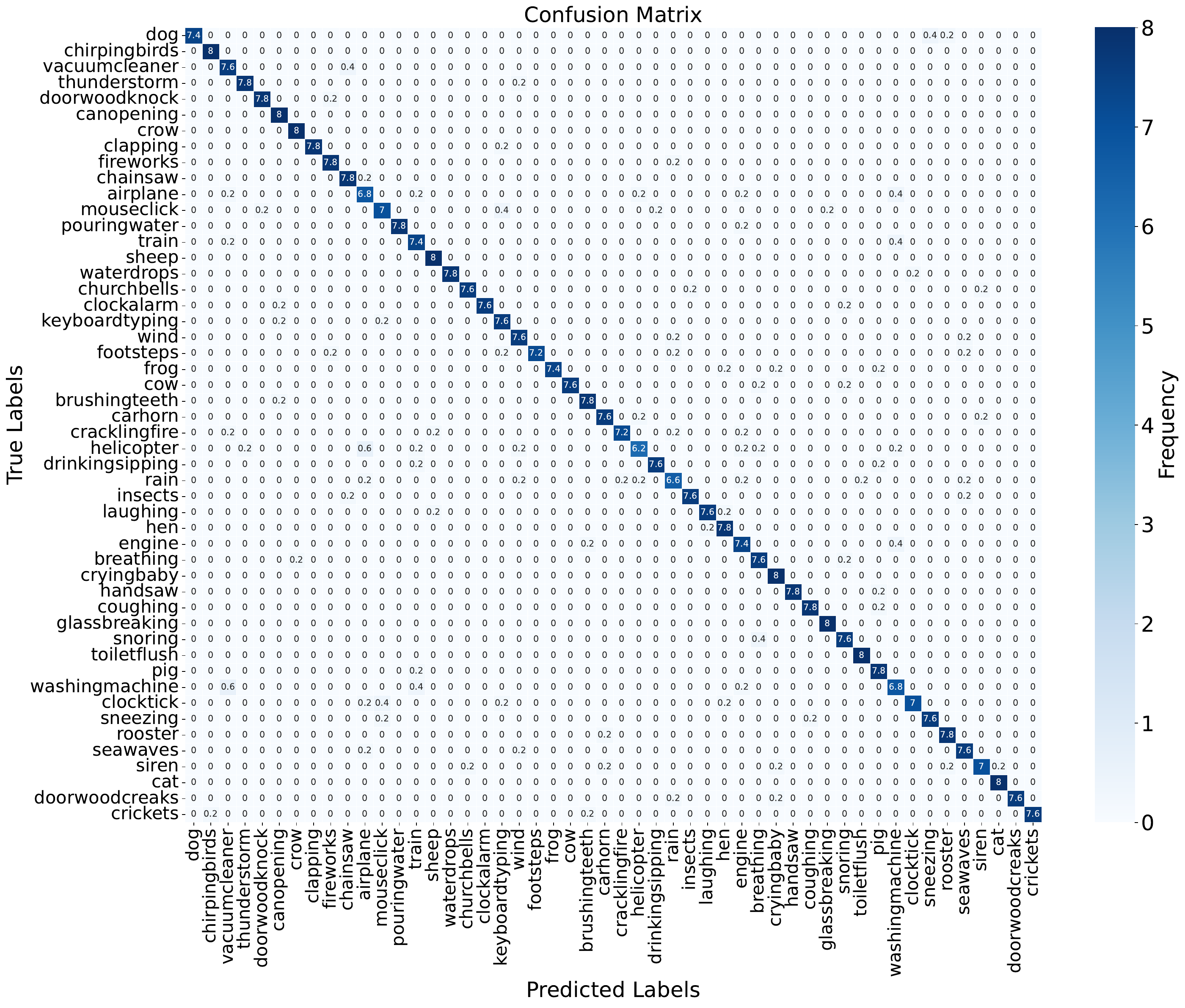}
        \caption{Averaged confusion matrix for QCNN14 ($p = 0.75$).}
        \label{fig:ESC-50_qcnn75_confusion}
    \end{subfigure}

    \caption{ESC-50 dataset: (a) convergence curve   and (b, c)  average of confusion matrices (CM) over 5-folds. Each fold-wise CM is obtained from QCNN14, where it gives maximum accuracy.}
\end{figure}

\begin{figure}[ht]
    \centering
    \begin{subfigure}{0.49\textwidth}
        \centering
        \includegraphics[scale=0.25]{Figures/average_convergence_QResNet38.pdf}
        \caption{Convergence curve averaged across 5-folds for ESC-50.}
        \label{fig:ESC-50_qresnet38_convergence}
    \end{subfigure}
    \hfill
    \begin{subfigure}{0.49\textwidth}
        \centering
        \includegraphics[scale=0.15]{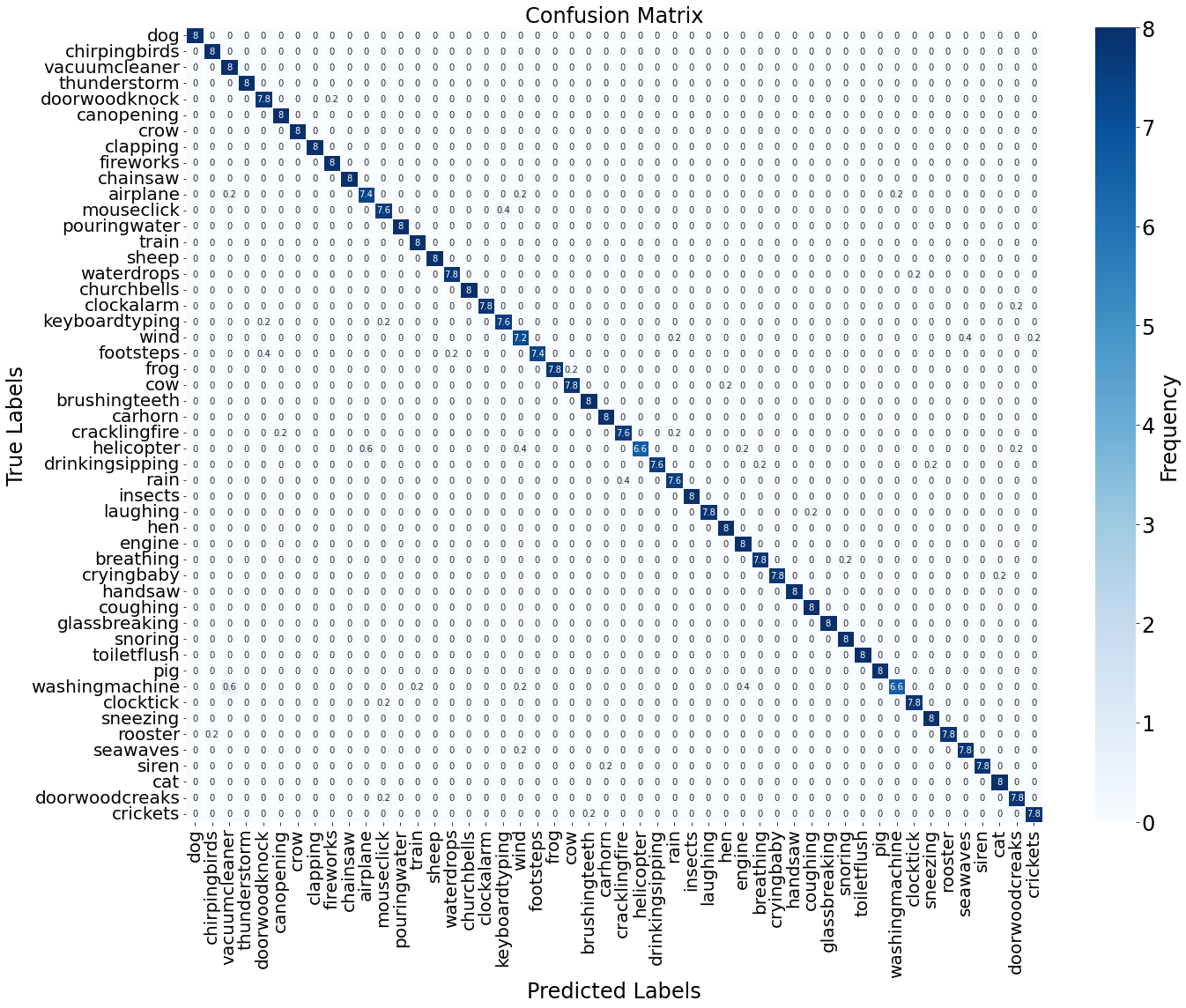}
        \caption{Averaged confusion matrix for QResNet38 ($p = 0.25$).)}
        \label{fig:ESC-50_qresnet38_confusion}
    \end{subfigure}

    \hfill
    \begin{subfigure}{0.48\textwidth}
        \centering
        \includegraphics[scale=0.15]{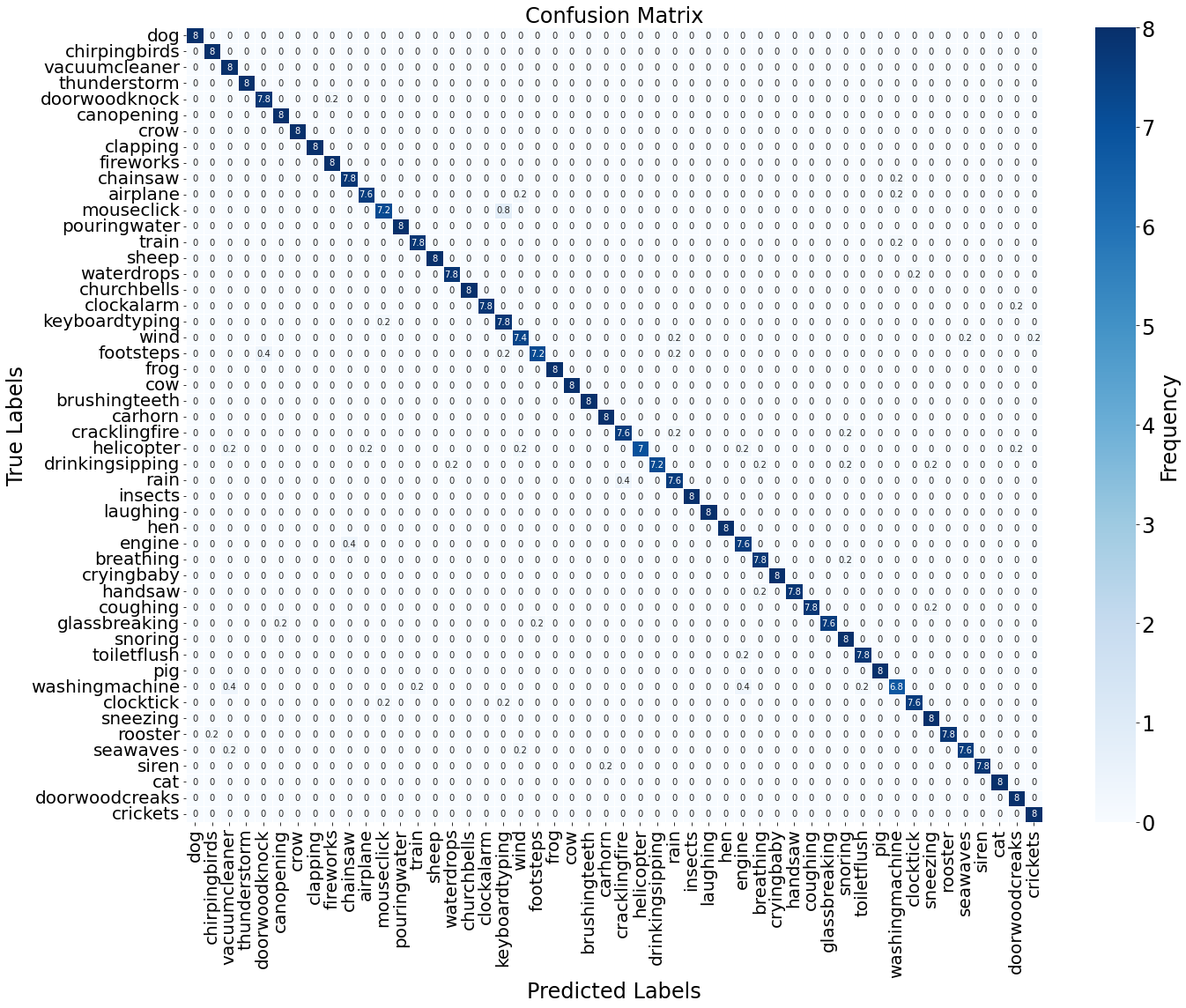}
        \caption{Averaged confusion matrix for QResNet38 ($p = 0.75$).}
        \label{fig:ESC-50_qresnet38_confusion}
    \end{subfigure}

    \caption{ESC-50 dataset: (a) convergence curve   and (b, c)  average of confusion matrices (CM) over 5-folds. Each fold-wise CM is obtained from QResNet38, where it gives maximum accuracy.}
\end{figure}



\begin{figure}[ht]
    \centering
    \begin{subfigure}{0.48\textwidth}
        \centering
        \includegraphics[scale=0.25]{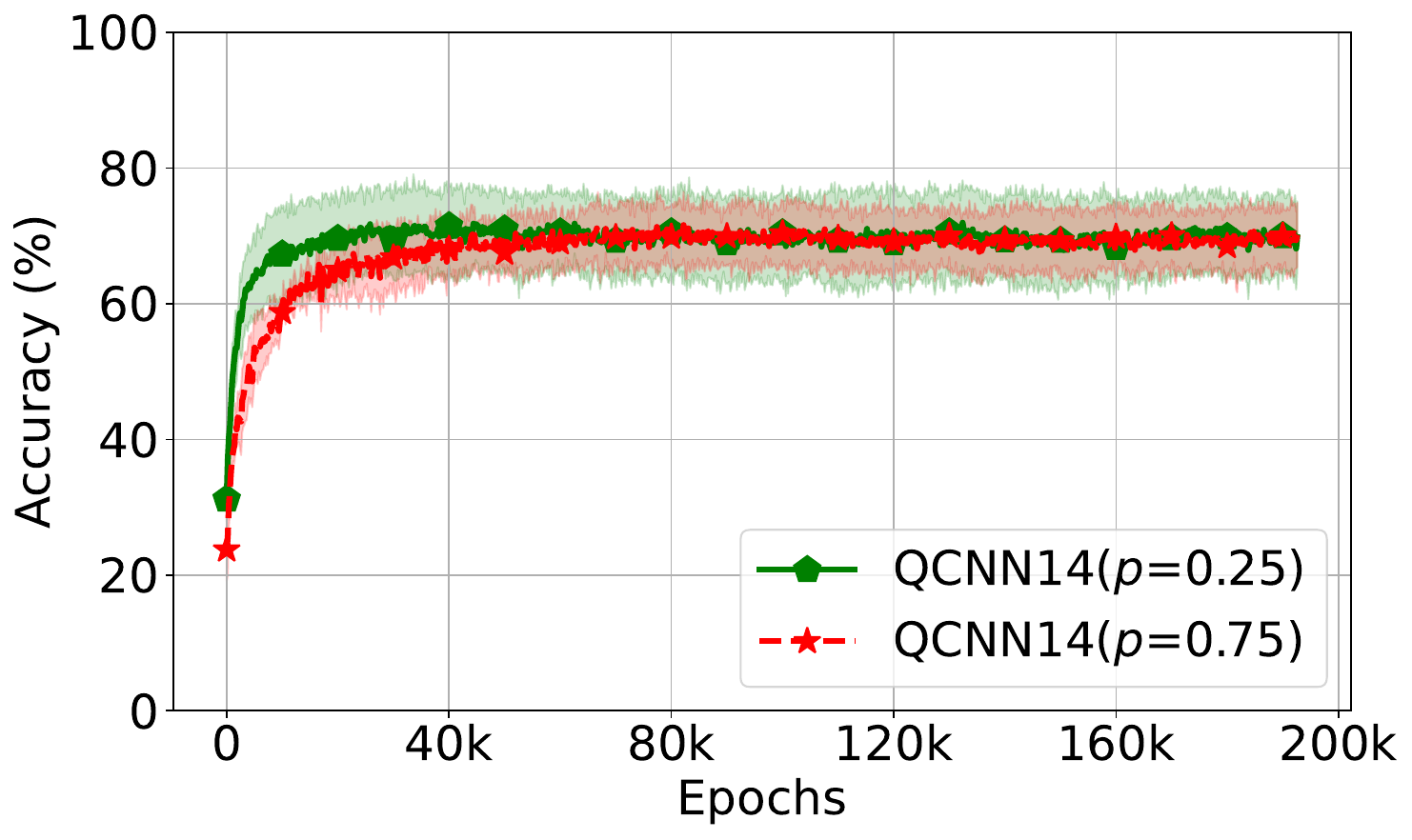}
        \caption{Convergence curve averaged across 4-folds for RAVDESS.}
        \label{fig:qcnn25_RAVDESS_convergence}
    \end{subfigure}
    \hfill
    \begin{subfigure}{0.48\textwidth}
        \centering
        \includegraphics[scale=0.25]{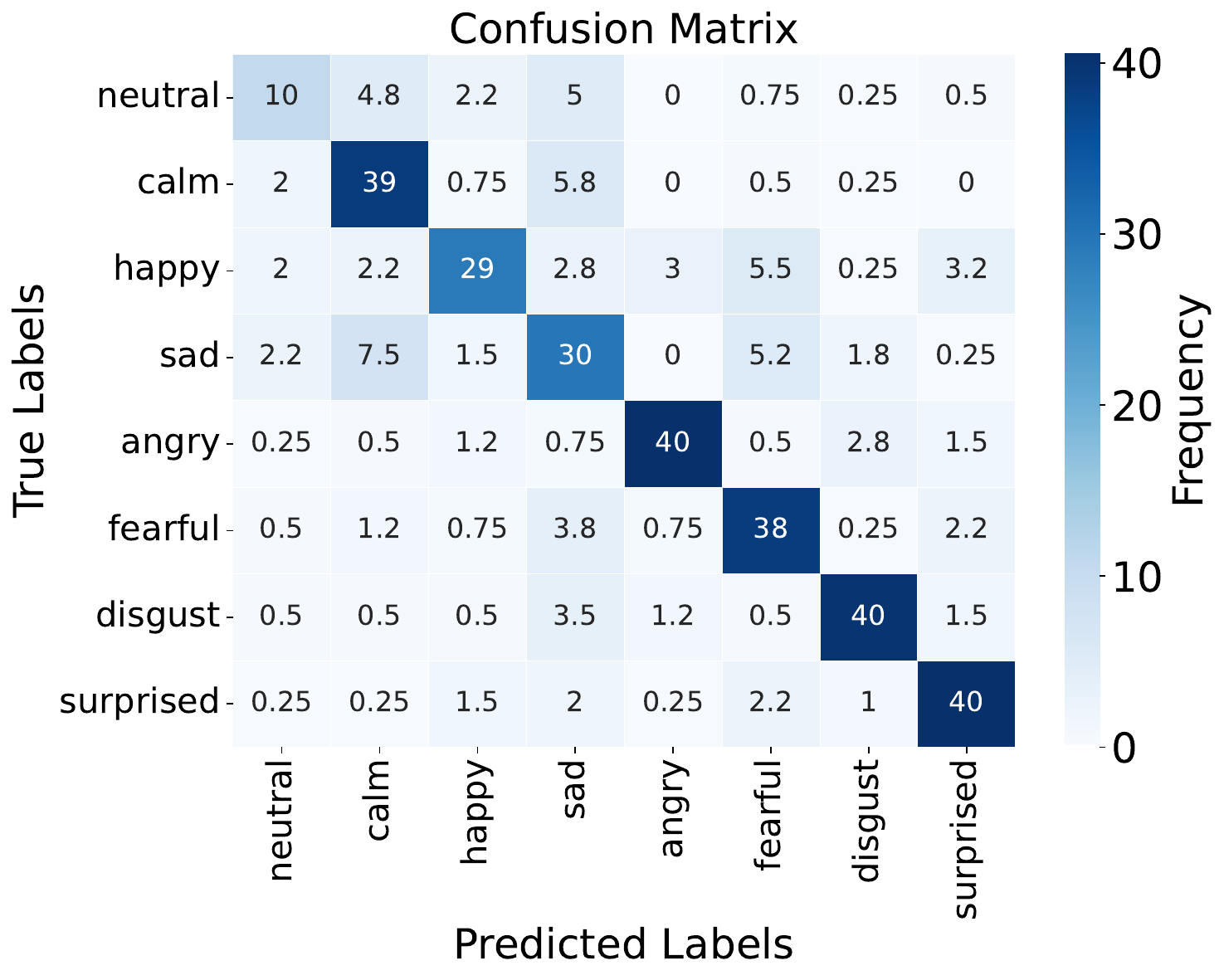}
        \caption{Averaged confusion matrix for QCNN14 ($p = 0.25$).}
        \label{fig:qcnn25_RAVDESS_confusion}
    \end{subfigure}

    \vspace{0.5cm} 

    \hfill
    \begin{subfigure}{0.48\textwidth}
        \centering
        \includegraphics[scale=0.25]{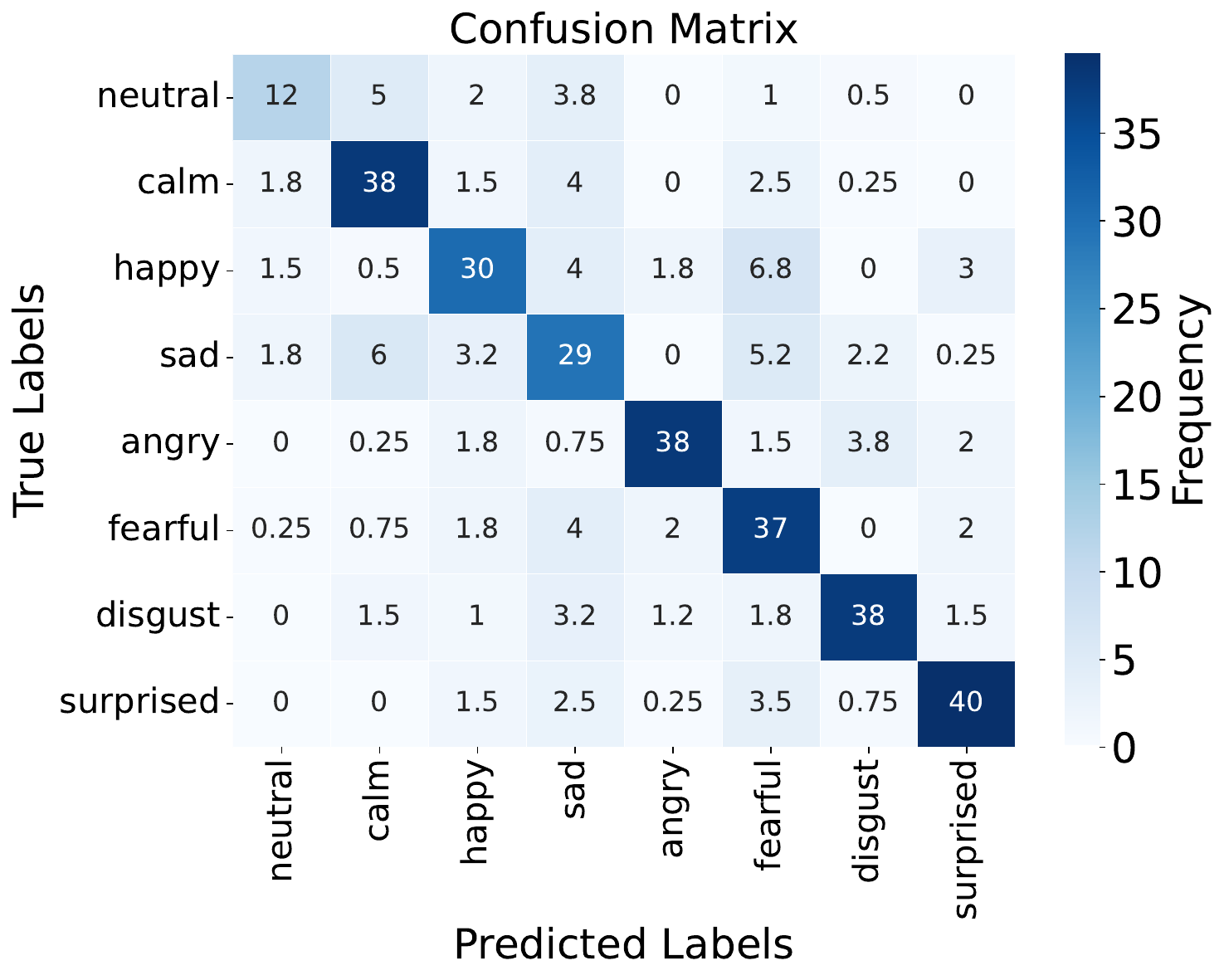}
        \caption{Averaged confusion matrix for QCNN14 ($p = 0.75$).}
        \label{fig:qcnn75_RAVDESS_confusion}
    \end{subfigure}

    \caption{RAVDESS dataset: (a) convergence curve   and (b, c)  average of confusion matrices (CM) over 4-folds. Each fold-wise CM is obtained from QCNN14, where it gives maximum accuracy.}
\end{figure}


\begin{figure}[ht]
    \centering
    \begin{subfigure}{0.48\textwidth}
        \centering
        \includegraphics[scale=0.25]{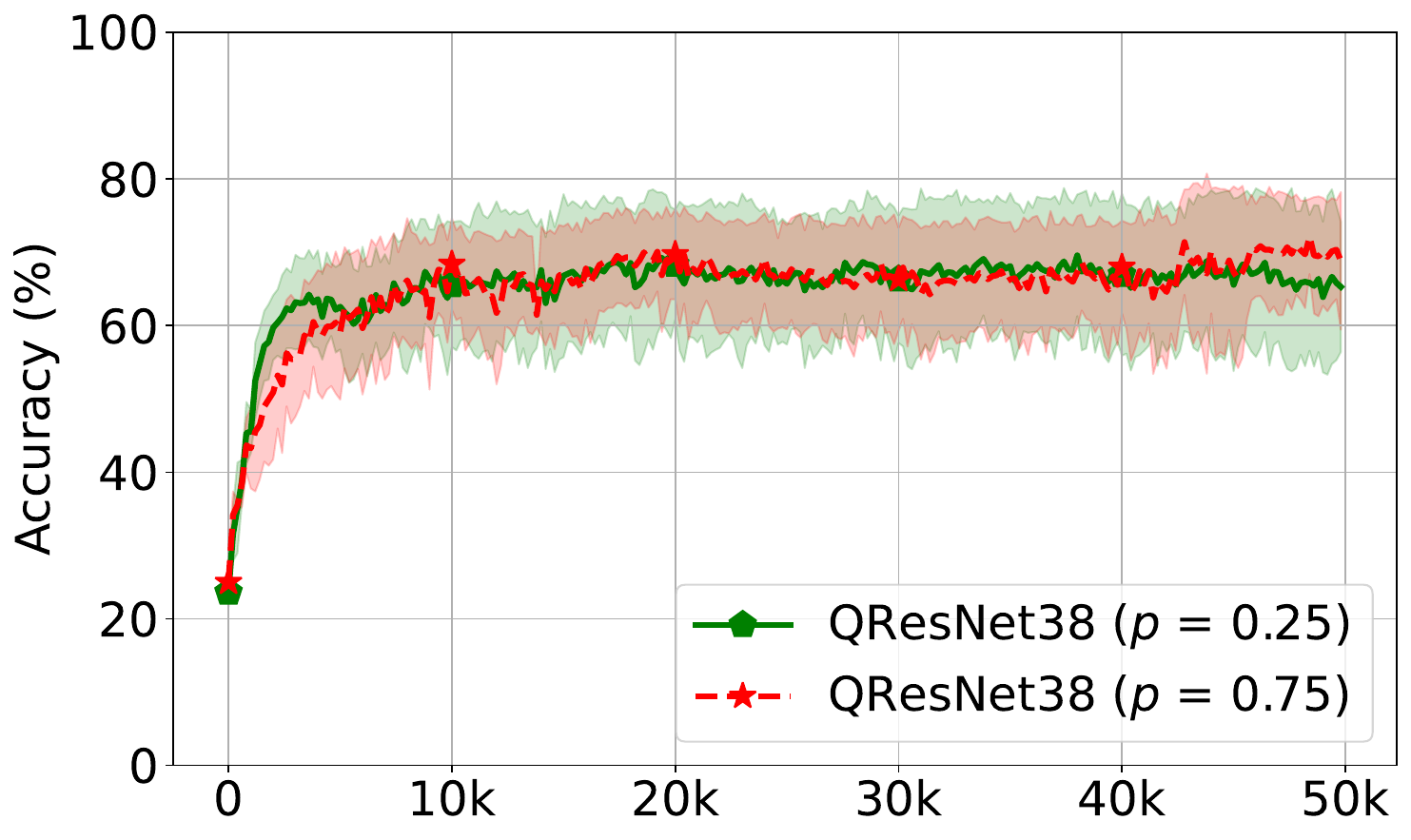}
        \caption{Convergence curve averaged across 4-folds for RAVDESS.}
        \label{fig:qcnn25_RAVDESS_convergence}
    \end{subfigure}
    \hfill
    \begin{subfigure}{0.48\textwidth}
        \centering
        \includegraphics[scale=0.25]{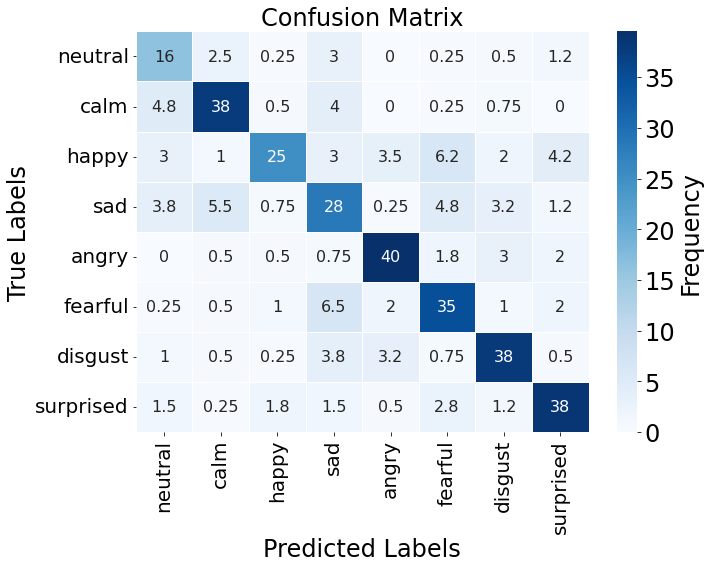}
        \caption{Averaged confusion matrix for QResNet38 ($p = 0.25$).}
        \label{fig:qcnn25_RAVDESS_confusion}
    \end{subfigure}

    \vspace{0.5cm} 

    \hfill
    \begin{subfigure}{0.48\textwidth}
        \centering
        \includegraphics[scale=0.25]{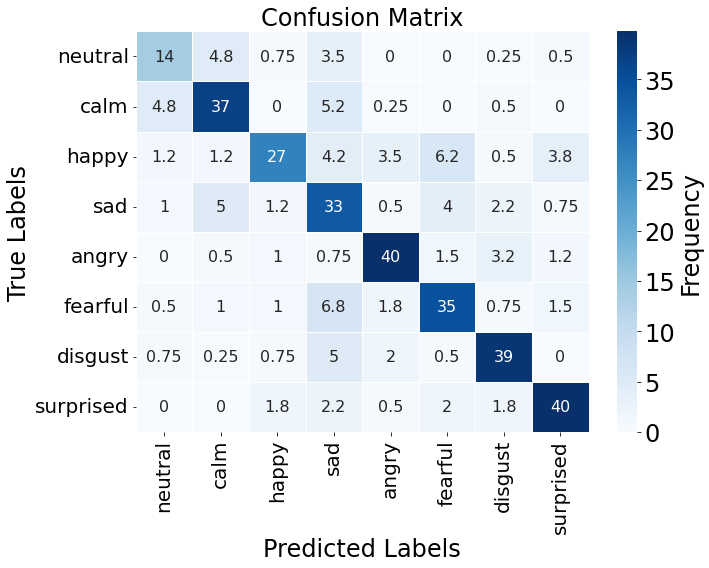}
        \caption{Averaged confusion matrix for QResNet38 ($p = 0.75$).}
        \label{fig:qcnn75_RAVDESS_confusion}
    \end{subfigure}

    \caption{RAVDESS dataset: (a) convergence curve   and (b, c)  average of confusion matrices (CM) over 4-folds. Each fold-wise CM is obtained from QResNet38, where it gives maximum accuracy.}
\end{figure}

\clearpage

\bibliographystyle{IEEEtran}
\bibliography{ref_tidy.bib}
\end{document}